\documentclass[12pt]{article}      

\usepackage{graphicx}
\usepackage{amssymb}
\usepackage{amsmath}
\usepackage{lineno}
\usepackage{textcomp}
\usepackage{float}
\usepackage{fancyvrb}
\usepackage{caption}
\usepackage{graphicx,amsmath,amsfonts,amsthm,bbm,setspace,url}
\usepackage{natbib,verbatim,mathrsfs}
\usepackage{subcaption}
\usepackage{floatpag}
\usepackage{numprint}
\npthousandsep{,}

\DeclareMathOperator*{\argmin}{arg\,min}

\advance\voffset by -0.25cm
\begin{document}

%\sloppy
\baselineskip=20pt

\begin{center}{
  \Large \bf
Modeling massive highly-multivariate nonstationary spatial data with the basis graphical lasso}

\bigskip
\bigskip

{\bf 
  Mitchell Krock\footnote[1]{Department of Statistics,
  Rutgers University. \\
  \quad \quad Corresponding author e-mail:
 \texttt{mk1867@stat.rutgers.edu}},
  %Author e-mail:  
  William Kleiber\footnote[2]{Department of Applied Mathematics,
  University of Colorado Boulder}, 
  Dorit Hammerling\footnote[3]{Department of Applied Mathematics and Statistics, Colorado School of Mines},
  and 
  Stephen Becker\footnotemark[2]
}

\bigskip
\bigskip

{\bf \today}

\end{center}

\bigskip
\bigskip

\begin{abstract}
We propose a new modeling framework for highly-multivariate spatial processes that synthesizes ideas from recent multiscale and spectral approaches with graphical models. The basis graphical lasso writes a univariate Gaussian process as a linear combination of basis functions weighted with entries of a Gaussian graphical vector whose graph is estimated from optimizing an $\ell_1$ penalized likelihood. This paper extends the setting to a multivariate Gaussian process where the basis functions are weighted with Gaussian graphical vectors. We motivate a model where the basis functions represent different levels of resolution and the graphical vectors for each level are assumed to be independent. Using an orthogonal basis grants linear complexity and memory usage in the number of spatial locations, the number of basis functions, and the number of realizations. An additional fusion penalty encourages a parsimonious conditional independence structure in the multilevel graphical model. We illustrate our method on a large climate ensemble from the National Center for Atmospheric Research's Community Atmosphere Model that involves 40 spatial processes.
\bigskip

\end{abstract}

%%%%%%%%%%%%%%%%%%%%%%%%%%%%%%%%%%%%%%%%%%%%%%%%%%%%%%%%%%%%%%%%%%%%%%%%%%%%%%%%
%%%%%%%%%%%%%%%%%%%%%%%%%%%%%%%%%%%%%%%%%%%%%%%%%%%%%%%%%%%%%%%%%%%%%%%%%%%%%%%%

\section{Introduction}

The past twenty years have seen a surge of interest in developing models for multivariate spatial processes. 
The major obstacle lies in defining valid cross-covariance functions that can characterize complex interactions between multiple processes; the primary difficulty is that the cross-covariance and marginal covariance functions must work together to provide a nonnegative definite matrix function. 
Most research has focused on exploring new models or new approaches for defining cross-covariances that are valid for a handful of processes. Many applied problems, such as those in statistical climatology, involve datasets with dozens to hundreds of variables, and existing approaches fail as strategies to model and understand relevant dependencies between variables.

To introduce the basic ideas, let us fix some notation. 
The $p$-variate observational Gaussian process model under consideration is  
\begin{equation}  \label{eq:obs.model}
\begin{pmatrix}
Y_1(\mathbf{s}) \\
\vdots \\
Y_p(\mathbf{s}) 
\end{pmatrix} 
= \begin{pmatrix} 
\mu_1(\mathbf{s}) \\
\vdots \\
\mu_p(\mathbf{s}) 
\end{pmatrix} 
+ 
 \begin{pmatrix} 
Z_1(\mathbf{s}) \\
\vdots \\
Z_p(\mathbf{s}) 
\end{pmatrix} 
+
 \begin{pmatrix} 
\varepsilon_1(\mathbf{s}) \\
\vdots \\
\varepsilon_p(\mathbf{s}) 
\end{pmatrix} \end{equation}
or in vector form,
$
\mathbf{Y}(\boldsymbol{s}) = \boldsymbol \mu(\boldsymbol{s}) + \mathbf{Z}(\boldsymbol{s}) + \boldsymbol \varepsilon(\boldsymbol{s})
$. Here, $\mathbf{Y}(\boldsymbol{s}) =  (Y_1(\mathbf{s}) , \dots, Y_p(\mathbf{s}))^{\mathrm{T}}$ is the observed process at location $\mathbf{s} \in \mathbb{R}^d$ with mean $\boldsymbol{\mu}(\mathbf{s}) =  (\mu_1(\mathbf{s}) , \dots, \mu_p(\mathbf{s}))^{\mathrm{T}}$ and spatially correlated stochastic variation $\mathbf{Z}(\mathbf{s}) =  (Z_1(\mathbf{s}) , \dots, Z_p(\mathbf{s}))^{\mathrm{T}}$, which we assume to be a multivariate Gaussian process. The observations are subject to noise $\boldsymbol{\varepsilon}(\mathbf{s}) = (\varepsilon_1(\mathbf{s}),\dots,\varepsilon_p(\mathbf{s}))^{\mathrm{T}}$, a mean zero multivariate white noise process with covariance matrix $\text{Cov}(\boldsymbol{\varepsilon}(\mathbf{s}),\boldsymbol{\varepsilon}(\mathbf{s})) = \text{diag}(\tau_1^2,\dots,\tau_p^2)$.

The paramount issue in working with multivariate processes is specifying the matrix-valued covariance, $\mathbf{C}(\mathbf{s}_1,\mathbf{s}_2) = (C_{ij}(\mathbf{s}_1,\mathbf{s}_2))_{i,j=1}^p$, where $C_{ij}(\mathbf{s}_1,\mathbf{s}_2) = \text{Cov}(Z_i(\mathbf{s}_1),Z_j(\mathbf{s}_2))$ are the direct and cross-covariance functions. This matrix function must be carefully constrained in order to be a nonnegative definite matrix function. In particular, for arbitrary locations $\{\mathbf{s}_i\}_{i=1}^n$, the block matrix $\boldsymbol \Sigma$ with $(i,j)$th block $\mathbf{C}(\mathbf{s}_i,\mathbf{s}_j)$ must be nonnegative definite.  \citet{MR3353096} give an overview of cross-covariance functions for multivariate geostatistics which is still relatively up-to-date. \citet{SALVANA2020100411} provide a more recent overview with a focus on multivariate spatio-temporal cross-covariance functions. The remainder of this paper considers Gaussian process models for continuous spatial processes with no temporal component. Spatial data with locations grouped by region (e.g.,\ by county or state) is known as areal or lattice data. We refer readers interested in high-dimensional multivariate spatio-temporal areal data to \citet{bradley2015, bradley2018}.

The main extant approaches to generating valid multivariate spatial models are reviewed by \citet{MR3353096} and include covariance convolution and kernel convolution \citep{gaspari1999,MR2324633,MR2682636,MR1631328}, the linear model of coregionalization  (LMC) \citep{goulard1992,schmidt2003,wackernagel2003multivariate, MR2154003}, latent dimensions \citep{MR2594414}, dynamical models \citep{MR1920802, MR2405328, MR2905058}, conditional Bayesian hierarchical structures \citep{MR2223933, MR2838443}, and direct specification as in the multivariate Mat\'ern \citep{MR2752612, MR2949350} and nonstationary extensions \citep{kleiber2012,kleiber2014}.  
\citet{cressie2016} develop a conditional model for multivariate spatial data. \citet{qadir2020} and \citet{qadir2021} work with the coherence function \citep{kleiber2017} to provide more flexible behavior than the multivariate Mat\'ern. \citet{vu2020} construct nonstationary and asymmetric covariances by embedding familiar covariances (e.g.,\ multivariate Mat\`ern) in a warped domain with warping functions obtained from deep learning. All of these models are designed to handle a few variables, typically fewer than five, and none are designed for high-dimensional multivariate spatial data.

To expand upon the previous statement, multivariate spatial data can be high-dimensional in the sense of a large number of observation locations ($n \gg 0)$ and/or a large number of output variables ($p \gg  0$).
Computational difficulties in the high-dimensional setting are unavoidable and particularly troublesome when both $n$ and $p$ are large.
Even with valid covariance and cross-covariance functions specified, which is already an issue if there are more than a handful of variables, likelihood computations and memory requirements for $p$-variate Gaussian processes at $n$ spatial locations scale as $\mathcal{O}(p^3 n^3)$ and $\mathcal{O}(p^2n^2)$, respectively. \citet{salvana2021} explore high-performance computing techniques to alleviate these costs. Modeling high-dimensional multivariate spatial processes requires specialized methodology, and most current techniques struggle when both $n$ and $p$ are large.

First, we describe some models for multivariate processes that focus on dealing with a large number of spatial locations $(n \gg 0)$. \citet{kleiber2019} generalize LatticeKrig \citep{nychka2015} to the multivariate case, relying on compactly supported basis functions and spatial autoregressive Gaussian Markov random field models for stochastic coefficients to handle massive spatial data. \citet{guinnessmultivar} is able to avoid maximum likelihood estimation and Bayesian inference by successively imputing data to an expanded lattice domain under a periodic model and efficiently estimating the cross-spectral density using Fast Fourier Transform techniques. 
 \citet{zhang_matrixnormal} perform Bayesian inference on massive multivariate spatial datasets by combining a matrix-normal distribution with Nearest Neighbor Gaussian Processes (NNGP) \citep{nngp}. Their model setup and choice of conjugate priors leads to closed form posterior distributions, and scalability is demonstrated on a dataset with over three million spatial locations. \citet{guinnessmultivar} and \citet{zhang_matrixnormal} consider a stationary bivariate setting in their data analyses, although these methods seem in principle able to scale to more than two variables.
 
Next, we turn attention to the ``highly-multivariate'' problem, where the number of variables $p$ is large. In one of the earliest efforts to tackle this problem, \citet{MR2836410} introduce ``aggregation cokriging'' for the prediction (but not the modeling) of highly-multivariate spatial processes.  \citet{dey2020graphical} propose a ``stitching'' of univariate Gaussian processes which preserves the marginal behavior of each univariate process and also conditional independencies between variables implied by a Gaussian graphical model. Stitching significantly reduces computational costs associated with highly-multivariate data, especially when used in  conjunction with a ``decomposable'' sparse graphical model, and it is also amenable to parallel computing.

%Finally, we note that several adaptations of LMC \citep{guinnessmultivar,LMCLidar,LMCortho, zhang_spatialfactor} are capable of handling data where both the number of variables and observation locations are large.

Finally, we note that several adaptations of LMC are capable of handling data where both the number of variables and observation locations are large. \citet{LMCLidar} use NNGP to model the latent process of LMC in a two-stage Bayesian hierarchical model, which they use to connect LiDAR maps with forest measurements. \citet{LMCortho} combine a sufficient statistic of the data and an orthogonal basis/loading matrix for substantial computational gains. \citet{zhang_spatialfactor} also develop a Bayesian LMC factor model with NNGP in the latent process, but they propose matrix-normal prior distributions and are able to avoid some constraints on the loading matrices used in \citet{LMCLidar}. \citet{meng2021b} build upon \citet{meng2021a}, where the loading matrix is stochastic with a Gaussian process prior on its elements, and employ variational inference \citep{titsias2010} for efficient computations. We discuss LMC in more detail in Section \ref{BGLmethodology}, as our formulation can be viewed as the opposite of LMC.  

Despite these advances, no one has truly addressed the problem of modeling complicated dependencies in space and across variables when both the number of variables and observed spatial locations are both large. 
The approaches in the previous paragraphs are all designed for modeling stationary processes, except for \citet{meng2021a} and \citet{meng2021b}, who do not attempt to model spatial data. Many datasets (e.g., our example from statistical climatology) exhibit strong nonstationarity both within and between processes, but models for nonstationary multivariate processes are not typically built to handle high-dimensional response vectors.  
Another issue is the ability to model multivariate processes on the globe, a task for which nonstationary covariance and cross-covariance functions are necessary \citep{jun2011}. 
This paper fills a major gap in the current literature by introducing methodology for highly-multivariate processes observed at a large number of spatial locations with nonstationary spatial and inter-variable dependencies.

We present a model for highly-multivariate and nonstationary spatial data that also accommodates estimation and simulation strategies for large networks of observation locations.  
The essential ideas rely on representing the vector-valued process in a basis expansion with sparsity-inducing Gaussian graphical modeling of the stochastic coefficients. We propose a penalized likelihood framework for estimation and associated optimization algorithms. 
 In addition to nonstationarity granted by judicious choice of basis functions and stochastic coefficient structure, employing {\it orthogonal} basis functions allows for rapid computation with non-gridded high-dimensional multivariate spatial data on an arbitrary domain.
The method is illustrated on a challenging climate data science problem involving $p=40$ spatial processes from an atmospheric model at thousands of locations over the globe that exhibit strong nonstationarities and cross-process dependencies. 
Our model provides straightforward interpretations of cross-process dependencies, which for the climate data example identify scientifically meaningful and justifiable relationships.

\section{Methodology} 

Our approach relies on a basis expansion of the spatially-correlated components of $\mathbf{Z}(\mathbf{s})$:
\begin{equation}\label{eq:multivarmodel}
  Z_i(s) = \sum_{\ell=1}^L W_{i\ell}  \phi_\ell(s) \quad (i=1,\dots,p)
\end{equation}
for some classes of basis functions $\{\phi_\ell\}$ and stochastic coefficients $\{W_{i \ell} \}$. 
For fixed $i$, this representation subsumes many popular approaches in the spatial statistical literature that have been primarily explored in the univariate setting, including discretized spectral methods, low rank approaches, and empirical orthogonal functions among others, despite potential limitations depending upon the choice of basis function \citep{stein2014}. 
Few extensions to the multivariate case (\ref{eq:multivarmodel}) have been made, with 
\citet{kleiber2019} being a notable case. In  Section \ref{BGLmethodology}, we show how the univariate basis graphical lasso directly extends to the multivariate setting (\ref{eq:multivarmodel}), and in Section \ref{blockdiagBGL}, we discuss an alternative which is more appropriate for massive multivariate spatial data.

\subsection{Basis Graphical Lasso}  \label{BGLmethodology}

Our modeling and optimization strategy follows from extensions to our prior work, which we discuss and connect to the multivariate problem in this section. 
In \citet{BGL}, we introduced the basis graphical lasso (BGL) to model (\ref{eq:multivarmodel}) for the univariate case $p=1$. 
The goal of the BGL is to obtain a sparse nonparametric estimate for the precision matrix $Q$ of the mean zero Gaussian graphical vector $\mathbf{W} = (W_1,\dots,W_L)^{\mathrm{T}}$ when
$Z(\mathbf{s}) = \sum_{\ell=1}^L W_\ell \phi_\ell(\mathbf{s})$.
In other words, we build a Gaussian process by fitting a Gaussian graphical model to the random coefficients of fixed basis functions. Recall that the sparsity pattern of the precision matrix encodes conditional independencies between random variables, with $Q_{ij} =0$ if and only if $W_i$ and $W_j$ are conditionally independent given all other entries of $\mathbf{W}$. This information is commonly visualized with an undirected graph known as a Gaussian graphical model, where vertices symbolize variables and the lack of an edge between two vertices indicates such a conditional independence \citep{rue2005}. 
Inspired by the graphical lasso \citep{friedman_sparse_2008}, we use an $\ell_1$ penalized likelihood to estimate the graphical model, but in the standard graphical lasso setting $\mathbf{W}$ is observed directly, whereas our model includes basis functions and noise. 
Our BGL method is viable for a very large sample size ($n$), and multiple realizations ($m$) are preferred but not required. 
Note that this formulation is opposite to LMC, which is not typically used in a univariate setting but estimates the deterministic weights of individual Gaussian processes.  
We claim that %, with careful and judicious choices of the penalty format,    NOTE I think that this can be deleted, the penalty is not a big issue for the generalization
the BGL generalizes to the multivariate setting and also that the optimization routine enjoys a similar computational framework.  

With $\mathbf{W}_\ell = (W_{1\ell},\dots,W_{p \ell})^{\mathrm{T}} \sim N(\mathbf{0},\mathbf{Q}_\ell^{-1})$, the multivariate basis model (\ref{eq:multivarmodel}) alternatively can be written 
\begin{equation}\label{eq:multivarvecmodel}
  \mathbf{Z}(\mathbf{s}) = \sum_{\ell=1}^L \phi_\ell(\mathbf{s}) \mathbf{W}_\ell
\end{equation}
so that each basis function is weighted with a $p$-variate random vector. 
Assuming the weight vectors are independent means that this model amounts to characterizing the inverse covariance matrices $\{\mathbf{Q}_1,\dots,\mathbf{Q}_L\}$.
In contrast, the standard LMC considers (\ref{eq:multivarvecmodel}) where $\{\phi_1(\mathbf{s}),\dots,\phi_L(\mathbf{s})\}$ are independent Gaussian processes with deterministic weights $\{ \mathbf{W}_1,\dots,\mathbf{W}_L \}$. 
We note that recent variants of LMC \citep{LMCLidar,LMCortho, zhang_spatialfactor, meng2021b} can handle a large number of variables, and in fact orthogonal basis functions are also exploited for computational gains in \citet{LMCortho}, but with a large number of spatial locations, such techniques ultimately amount to efficiently modeling spatially-dense univariate Gaussian processes (e.g., \cite{nngp,variationallearning}). The semiparametric latent factor model \citep{Teh2005SemiparametricLF} shares a setup similar to LMC where a multi-output Gaussian process is represented as a linear mixture of independent univariate Gaussian processes. With these models, covariance kernel parameters of the independent latent Gaussian processes and the weight vectors (i.e.,\ columns of the loading matrix) must be estimated. Typically, the covariance kernels are stationary and the loading matrix is deterministic. Some nonstationary versions of LMC \citep{MR2154003, meng2021b} consider a stochastic loading matrix that depends on the input domain, which entails a prior distribution and a challenging Bayesian framework. To our knowledge, such models have not yet been tested with high-dimensional nonstationary spatial data.
On the other hand, our approach associates each basis function with a $p$-variate Gaussian graphical model, which is much faster from a computational point of view and automatically produces nonstationary covariance and cross-covariance functions with straightforward interpretations.
    
Let us assume $\boldsymbol \mu \equiv 0$ in (\ref{eq:obs.model}) for simplicity of exposition. 
In a typical mean function regression context, we can use generalized least squares and profiled likelihoods to estimate the regression coefficients. 
Given data at locations $\mathbf{s}_1,\dots,\mathbf{s}_n$, form the observation vector $\mathbf{Y} = (\mathbf{Y}(\mathbf{s}_1)^{\mathrm{T}},\dots,\mathbf{Y}(\mathbf{s}_n)^{\mathrm{T}})^{\mathrm{T}}$. 
Suppose we have multiple independent realizations $\mathbf{Y}_1,\dots,\mathbf{Y}_m$ of $\mathbf{Y}$. 
Note that all methodology developed in this paper can work with $m=1$ realization, but the task of learning the precision matrix of the basis weight vector is better suited for a setting with multiple realizations. Up to multiplicative and additive constants not depending on the $n p \times np$ variance-covariance matrix  $\boldsymbol  \Sigma = \text{Var}(\mathbf{Y})$, the negative log-likelihood is
\begin{equation}  \label{Sigma.likelihood.sum}
  \log \det \boldsymbol \Sigma + \frac{1}{m} \sum_{i=1}^m 
  \mathbf{Y}_i^{\mathrm{T}}\boldsymbol \Sigma^{-1} \mathbf{Y}_i.
\end{equation}
Define the sample covariance $\mathbf{S} = \frac{1}{m} \sum_{i=1}^m \mathbf{Y}_i \mathbf{Y}_i^{\mathrm{T}}$. 
Using the cyclic property of trace, we rewrite the negative log-likelihood as 
\begin{equation}  \label{Sigma.likelihood.S}
  \log \det \boldsymbol \Sigma + \text{tr}(\mathbf{S}  \boldsymbol \Sigma^{-1})
\end{equation}
to align with more prevalent notation in graphical lasso literature. 
Note that $\mathbf{Y}$ is simply a linear combination of the random coefficient vector $\mathbf{W} = (\mathbf{W}_1^{\mathrm{T}},\dots,\mathbf{W}_L^{\mathrm{T}})^{\mathrm{T}}$ and basis functions, plus noise. 
That is, $\mathbf{Y} = \boldsymbol \Phi \mathbf{W} + \boldsymbol \varepsilon$ for a basis matrix $\boldsymbol \Phi$, so $ \boldsymbol \Sigma = \boldsymbol \Phi \mathbf{Q}^{-1} \boldsymbol \Phi^{\mathrm{T}} + \mathbf{D}$. 
These matrices are defined explicitly in Section \ref{implement}, but it is important to realize here that the matrix algebra produces the same optimization problem as in the univariate case.

The original BGL solves the $\ell_1$-penalized maximum likelihood equation 
\begin{equation}  \label{BGL}
 \hat{ \mathbf Q} \in \argmin \limits_{\boldsymbol Q \succeq 0}  \  
  \log \det( \boldsymbol \Phi \mathbf{Q}^{-1} \boldsymbol \Phi^{\mathrm{T}} + \mathbf{D}) 
  + \text{tr}( \mathbf{S} (  \boldsymbol \Phi \mathbf{Q}^{-1} \boldsymbol \Phi^{\mathrm{T}} + \mathbf{D} )^{-1}) 
  +  \| \Lambda \circ \mathbf{Q}\|_1.
\end{equation}
We use the notation $\mathbf Q \succeq 0$ to indicate that $\mathbf Q$ is positive semidefinite. 
Here, $ \| \Lambda \circ \mathbf{Q}\|_1 =  \sum_{i,j} \Lambda_{ij} |\mathbf{Q}_{ij}|$ where $\Lambda_{ij}$ are nonnegative penalty parameters that  encourage sparsity in the estimate. 
Evaluating (\ref{BGL}) requires an expensive $\mathcal{O}(p^3n^3)$ Cholesky decomposition.  
After applying the matrix determinant lemma, the Sherman-Morrison-Woodbury formula, and the cyclic property of trace, we can equivalently minimize
\begin{equation}  \label{Qlikelihood.pl}
    \log \det \left( \mathbf{Q}  + \boldsymbol \Phi^{\mathrm{T}} \mathbf D^{-1} \boldsymbol \Phi \right) - \log \det \mathbf{Q}
- \text{tr} \left(  \boldsymbol \Phi^{\mathrm{T}} \mathbf D^{-1} \mathbf{S} \mathbf D^{-1} \boldsymbol \Phi  ( \mathbf{Q} + \boldsymbol \Phi^{\mathrm{T}} \mathbf D^{-1} \boldsymbol \Phi)^{-1}  \right)     + \|\Lambda \circ \mathbf Q\|_1.
\end{equation}
Once the matrices $\boldsymbol \Phi^{\mathrm{T}} \mathbf{D}^{-1} \boldsymbol \Phi$ and $ \boldsymbol \Phi^{\mathrm{T}} \mathbf D^{-1} \mathbf{S} \mathbf D^{-1} \boldsymbol \Phi $ are computed, evaluating (\ref{Qlikelihood.pl}) only requires Cholesky decompositions in the dimension $p L$, so we can reduce likelihood evaluations to $\mathcal{O}(p^3 L^3)$. 
However, (\ref{Qlikelihood.pl}) is nonsmooth and nonconvex with respect to $\mathbf{Q}$, so the minimization problem is nontrivial.

Studying the convexity/concavity structure of (\ref{Qlikelihood.pl}) suggests a difference-of-convex (DC) algorithm where the next guess for $\mathbf{Q}$ is obtained by solving a convex optimization problem with the concave part linearized at the previous guess. 
Such an algorithm reads 
\begin{equation}\label{QUICproblem}
  \mathbf{Q}^{(j+1)} = \argmin_{\mathbf{Q} \succeq 0}   \left(  - \log \det \mathbf{Q}+\text{tr} 
  \left(  \boldsymbol \Psi ^{(j)} \mathbf Q\right) + \|\Lambda \circ \mathbf Q\|_1 \right)
\end{equation}
where the linearization matrix
\begin{equation} \label{linearization}
\boldsymbol \Psi ^{(j)} 
= (\mathbf{Q}^{(j)} + \boldsymbol \Phi^{\mathrm{T}} \mathbf D^{-1} \boldsymbol \Phi )^{-1}
 + (\mathbf{Q}^{(j)} + \boldsymbol \Phi^{\mathrm{T}} \mathbf D^{-1} \boldsymbol \Phi )^{-1} \boldsymbol \Phi^{\mathrm{T}} \mathbf D^{-1} \mathbf{S} \mathbf D^{-1} \boldsymbol \Phi 
(\mathbf{Q}^{(j)} + \boldsymbol \Phi^{\mathrm{T}} \mathbf D^{-1} \boldsymbol \Phi )^{-1} \\
%= (\mathbf{Q}^{(j)} + \boldsymbol \Phi^{\mathrm{T}} \mathbf D^{-1} \boldsymbol \Phi )^{-1}
%(\mathbf{Q}^{(j)} + \boldsymbol \Phi^{\mathrm{T}} \mathbf D^{-1} \boldsymbol \Phi + \boldsymbol \Phi^{\mathrm{T}} \mathbf D^{-1} \mathbf{S} \mathbf D^{-1} \boldsymbol \Phi )
%(\mathbf{Q}^{(j)} + \boldsymbol \Phi^{\mathrm{T}} \mathbf D^{-1} \boldsymbol \Phi )^{-1}
\end{equation} 
is a function of the previous guess $\mathbf{Q}^{(j)}$ and the aforementioned precomputed matrices. 
Since a DC algorithm such as (\ref{QUICproblem}) is a majorization-minimization algorithm, we are guaranteed that the guesses for $\mathbf Q$ create a nonincreasing sequence in the objective function (\ref{Qlikelihood.pl}). 
Moreover, (\ref{QUICproblem}) is a well-studied problem known as the graphical lasso. Typically, the graphical lasso uses the  sample covariance matrix of directly-observed, nonnoisy variables to produce a sparse inverse covariance matrix and accordingly a graphical model for the variables. Here we iteratively call the graphical lasso algorithm to estimate a graph for latent basis weights (with additive noise in the observational model), and the linearization matrix (\ref{linearization}) acts as the sample covariance in the algorithm.
We solve the graphical lasso with the second order method \texttt{QUIC} \citep{QUIC}.% which has an easy-to-use \texttt{R} package.
 
\subsection{Multivariate Basis Graphical Lasso} \label{blockdiagBGL}

Although the previous section shows that the BGL can be readily extended to the multivariate setting, the generalization is not well-motivated by a connection to standard multivariate spatial models, and moreover it will require burdensome matrix calculations in the dimension $pL$. 
In particular, the BGL must compute the linearization matrix (\ref{linearization}) and substitute it into the graphical lasso at each step of the DC algorithm. With our climate data example we use $L=2000$ basis functions and $p=40$ variables; an 80,000 dimensional precision matrix is too large for this procedure. Advances in graphical modeling \citep{lineartime2019} may allow for estimation of graphs of this magnitude, but storing the dense linearization matrix poses an issue to further scalability.
We conclude the paper with more commentary about this direct generalization (see Section \ref{conclusions}), but here we propose a similar DC algorithm which is more feasible in a highly-multivariate setting. 

Our basic model still follows a penalized likelihood-based framework, minimizing 
\begin{equation}  \label{m.BGL}
 \hat{ \mathbf Q} \in \argmin \limits_{\boldsymbol Q \succeq 0}  \  
  \log \det( \boldsymbol \Phi \mathbf{Q}^{-1} \boldsymbol \Phi^{\mathrm{T}} + \mathbf{D}) 
  + \text{tr}( \mathbf{S} (  \boldsymbol \Phi \mathbf{Q}^{-1} \boldsymbol \Phi^{\mathrm{T}} + \mathbf{D} )^{-1}) 
  + P(\mathbf{Q})
\end{equation}
for some penalty $P$.
However, an $\ell_1$ graphical lasso-type penalty by itself does not impose any regularity on the {\em structure} of coefficient graphs. 
To motivate our proposal, we recall some recent insights into multivariate modeling that will suggest an appropriate form for $P$.

The multivariate spectral representation theorem states 
\begin{equation}  \label{spectralintegral}
  \mathbf{Z}(\mathbf{s}) =  \int \exp(i \boldsymbol \omega^{\mathrm{T}} \mathbf{s}) \mathbf{W}(\mathrm{d} \boldsymbol \omega)
\end{equation} 
for a mean zero stationary process $\mathbf{Z}(\mathbf{s})$, where $\mathbf{W}(\cdot)$ is a complex-valued mean zero random measure vector \citep{stein1999}. 
Taking a discretization of the integral we approximate 
\begin{equation}  \label{spectralsum}
  \int \exp(i \boldsymbol \omega^{\mathrm{T}} \mathbf{s}) \mathbf{W}(\mathrm{d} \boldsymbol \omega) \approx \sum_\ell \cos(\boldsymbol \omega^{\mathrm{T}}_\ell \mathbf{s} + \theta_\ell) \mathbf{W}_\ell
\end{equation} 
to motivate writing (\ref{eq:multivarvecmodel}) as a linear combination of {\it independent} random vectors $\mathbf{W}_\ell$. 
As an aside, another justification for modeling the coefficient vectors $\mathbf{W}_\ell$ as independent across $\ell$ is the Karhunen-Lo\'eve expansion, in which the basis functions $\phi_\ell$ are eigenfunctions and the random coefficients are theoretically independent.  
Indeed, in our climate modeling example below we use a discrete approximation to the Karhunen-Lo\'eve expansion from which independence of coefficient vectors is expected.

The spectral representation theorem (\ref{spectralintegral}) is intimately linked to the spectral density matrix of $\mathbf{Z}(\mathbf{s})$ where we identify $\text{Var}(\mathbf{W}_\ell)$ with $\mathbf{f}(\boldsymbol \omega_\ell)$. 
One connection to our model is that $\mathbf{Q}_\ell^{-1}$ can be viewed as the spectral density matrix at frequency $\boldsymbol \omega_\ell$, but 
our approach estimates the {\em inverse} spectral density matrix in a Gaussian graphical framework, and moreover, we consider replacing the harmonic basis functions with other globally-supported multiresolution basis functions. 
Indeed, \citet{kleiber2017} provides interpretation and exploration of spectral coherence that will additionally motivate our penalized likelihood implementation. 
%We continue the theme of estimating the sparse inverse covariance matrix $\mathbf{Q}_\ell$ rather than $\mathbf{Q}_\ell^{-1}$. 
Before moving on, it is important to note that although we use the spectral representation theorem (\ref{spectralintegral}) to motivate the ensuing approach, our method is general and extends beyond harmonic basis expansions but with similar coherence-like interpretations of coefficient dependence. 
\cite{guinnessmultivar} proposes a multivariate space-time model with flexible coherence structures which uses LMC in the spectral domain, but the covariance structure is stationary and the method is only viable for gridded data. 
We remind the reader that LMC also relies on an independence assumption where $\mathbf{Z}(\mathbf{s})$ is a linear combination of independent univariate Gaussian processes.

The multivariate basis graphical lasso model can be motivated with the same penalized likelihood context as in Section \ref{BGLmethodology}. 
Recall the model setup: we work under the additive model (\ref{eq:obs.model}) with $\mathbf{Z}(\mathbf{s})$ specified as in the basis representation (\ref{eq:multivarvecmodel}), and $\mathbf{W}_i$ and $\mathbf{W}_j$ are independent Gaussian graphical vectors for $i\neq j$.  
In particular, we assume that $\mathbf{Q}_\ell = \text{Var}(\mathbf{W}_\ell)^{-1}$ is a sparse matrix defining a graphical structure at level $\ell$. 
%To estimate the precision matrix entries governing the graphical structure, we propose a penalized likelihood framework that minimizes the negative log-likelihood plus a penalty $P(\mathbf{Q}_1,\dots,\mathbf{Q}_L).$
%\begin{equation}  \label{eq:pen.lik}
%\log \det \boldsymbol \Sigma + \text{tr}(\mathbf{S}  \boldsymbol \Sigma^{-1}) + P(\mathbf{Q}_1,\dots,\mathbf{Q}_L).
%\end{equation}
%Different choices of $P$ will have different implications on the assumed process structure. 
If we consider each $\mathbf{Q}_\ell$ to correspond to an arbitrary sparse graphical model, then we propose $P$ as a graphical lasso regularization for each level:
\begin{equation} \label{unfusedpenalty}
P(\mathbf{Q}_1,\dots,\mathbf{Q}_L) =  
 \lambda \sum_{\ell=1}^L \sum_{i \neq j} \left\vert(\mathbf{Q}_{\ell})_{ij} \right\vert.
\end{equation}
This penalty enforces sparsity for each precision matrix  but not necessarily any similarity between levels of resolution. 
Recent development in spectral coherence \citep{kleiber2017} suggests that we should expect the coherence of processes arising in practice to vary smoothly across levels. 
In addition to the $\ell_1$ sparsity penalty, we include an $\ell_1$ sequentially-fused penalty to encourage similarity of the conditional independence structure across adjacent levels of resolution:
\begin{equation} \label{fusedpenalty}
  P(\mathbf{Q}_1,\dots,\mathbf{Q}_L) =  
 \lambda \sum_{\ell=1}^L \sum_{i \neq j} \left\vert(\mathbf{Q}_{\ell})_{ij} \right\vert
  + \rho  \sum_{\ell=1}^{L-1} \sum_{i \neq j} \left\vert (\mathbf{Q}_{\ell})_{ij} -(\mathbf{Q}_{\ell+1})_{ij} \right\vert.
\end{equation}
The new fusion penalty with tuning parameter $\rho$ penalizes precision matrices at adjacent levels if their off-diagonals do not have the same value. This formulation suggests a smoothly-varying graph structure and produces a parsimonious conditional independence structure of the random weights over all levels of resolution.
 
Assuming that $\mathbf{Q}= \text{diag}(\mathbf{Q}_1,\dots,\mathbf{Q}_L)$ does not change any reasoning leading to the BGL formulation from Section \ref{BGLmethodology} but allows us to reduce computations on matrices of size $pL \times pL$ to computations on $L$ $p \times p$ matrices. At each step of the DC algorithm (\ref{QUICproblem}), we simplify the minimization problem to
\begin{equation}\label{blockdiagpenaltyproblem}
 \argmin_{\mathbf{Q}_{\ell} \succeq 0, \ \ell=1,\dots,L}   \left(   \sum_{\ell=1}^L - \log \det \mathbf{Q}_\ell +\text{tr} 
  \left(  \boldsymbol \Psi_{\ell} \mathbf Q_\ell \right)  \right) + P(\mathbf{Q}_1,\dots,\mathbf{Q}_L)
\end{equation}
where $\boldsymbol \Psi_\ell$ is the $\ell$\textsuperscript{th} block diagonal of the linearization matrix $\boldsymbol\Psi$, which depends upon the previous graph guesses. %, and $P(\mathbf{Q}_1,\dots,\mathbf{Q}_L)$  denotes the penalty. 
An efficient way to calculate $\boldsymbol\Psi_1,\dots,\boldsymbol\Psi_L$ for orthogonal bases is presented in Section \ref{implement}. 

Using only the sparsity penalty (\ref{unfusedpenalty}) means (\ref{blockdiagpenaltyproblem}) separates into $L$ independent graphical lasso problems with ``sample covariance'' matrices $\boldsymbol\Psi_1,\dots,\boldsymbol\Psi_L$.\footnote{See Section \ref{initguess}.}
Using the fusion penalty (\ref{fusedpenalty}) gives no such separation of (\ref{blockdiagpenaltyproblem}) into $L$ independent optimization problems at each DC iteration. 
Instead, we treat $\boldsymbol\Psi_1,\dots,\boldsymbol\Psi_L$ as an array of sample covariance matrices and substitute them into the fused multiple graphical lasso (FMGL) \citep{MGL} for modeling multiple similar graphical models across multiple datasets. 
A related idea is the joint graphical lasso \citep{JGL}, but their similarity-inducing regularization term penalizes all pairs of graphs rather than just adjacent graphs as in (\ref{fusedpenalty}), suggesting behavior similar to white noise processes---an unreasonable assumption for most real spatial data applications. 
The associated algorithm uses a slower optimization approach which calculates eigendecompositions in \texttt{R}. 
The FMGL algorithm, written in Matlab, uses the same second order approximation as \texttt{QUIC} \citep{QUIC} and also exhibits quadratic convergence. 
A general framework for second order optimization in high-dimensional statistical modeling with regularization is developed in \cite{NIPS2014_5578}.

\section{Implementation Strategy}  \label{implement}

%\subsubsection{FBGL implementation} 

We highlight some important aspects of implementing the multivariate BGL model. In practice, we independently estimate an error variance for each variable and define the diagonal matrix $\mathbf{D}$ accordingly.\footnote{See Section  \ref{errorvariance_estimate}.}
Matrices $\boldsymbol \Phi^{\mathrm{T}} \mathbf{D}^{-1} \boldsymbol \Phi$ and $ \boldsymbol \Phi^{\mathrm{T}} \mathbf D^{-1} \mathbf{S} \mathbf D^{-1} \boldsymbol \Phi $ must be computed effectively---ignoring $\mathbf{D}^{-1}$, the naive matrix multiplications cost $\mathcal{O}(p^3 n L^2)$ and $\mathcal{O}(p^3 n^2 L )$, respectively. Moreover, the block diagonals of the linearization (\ref{linearization}) must be retrieved without expending $\mathcal{O}(p^3 L^3)$ flops for the matrix inverse $(\mathbf{Q} + \boldsymbol \Phi^{\mathrm{T}} \mathbf{D}^{-1} \boldsymbol \Phi)^{-1}$. In \cite{BGL} with $p=1$ this linearization step was not an issue---the computational bottleneck was iteratively solving the graphical lasso in the dimension of the basis functions $L$. Here we are modeling $L$ graphs of dimension $p$, so the graphical modeling step may no longer be the bottleneck  but rather the linearization (\ref{linearization}). However, assuming that $\mathbf{Q}$ is block diagonal and basis functions are orthogonal makes this linearization step trivial.

Let $\Phi$ be the $n \times L$ basis matrix with $(i,j)$ entry $\phi_j(\mathbf{s}_i)$. Observe that
\begin{equation} \label{introvec}
\mathbf{Z}(\mathbf{s}_i)
= \sum_{\ell=1}^L \phi_\ell(\mathbf{s}_i) \mathbf{W}_\ell
 =  \mathbf{M} \Phi^{\mathrm{T}}_i \quad \quad (i=1,\dots,n)
 \end{equation}
where $\mathbf{M}$ has columns $\mathbf{W}_1,\dots,\mathbf{W}_L$ and $\Phi^{\mathrm{T}}_i$ is the $i$\textsuperscript{th} column of $\Phi^{\mathrm{T}}$.
Introducing the vec$(\cdot)$ operator, which stacks the columns of a matrix one-by-one into a vector, we write the process observation vector as 
\begin{equation} \label{stackvec}
\mathbf{Z}=\begin{pmatrix}
\mathbf{Z}(\mathbf{s}_1)\\
\vdots \\
\mathbf{Z}(\mathbf{s}_n)
\end{pmatrix} 
= 
\begin{pmatrix}
\mathbf{M} \Phi_1^{\mathrm{T}}
\\ 
\vdots \\ 
\mathbf{M} \Phi_n^{\mathrm{T}}
\end{pmatrix} 
= \text{vec}(\mathbf{M} \Phi^{\mathrm{T}})
\end{equation}
where the last equation follows from (16.2.7) in \cite{MR1467237}. The vec operator cooperates with the Kronecker product $\otimes$ in the following way: $\text{vec}(ABC) = (C^{\mathrm{T}} \otimes A) \text{vec}(B)$ whenever $ABC$ is well-defined. For us, this implies
\begin{equation} \label{kroneckermodelZ}
\mathbf{Z} = (\Phi \otimes I_p) \mathbf{W}
\end{equation}
since $\mathbf{W} = \text{vec}(\mathbf{M})$  by construction. Thus we identify $\boldsymbol \Phi = \Phi \otimes I_p$. Also note that $\mathbf{D}^{-1} =I_n \otimes   \text{diag}(\tau_1^{-2},\dots,\tau_p^{-2})$ is a Kronecker product. Since $(A \otimes C) (B \otimes D) = AB \otimes CD$ whenever $AB$ and $CD$ are well-defined, we have the Kronecker product representation
$\boldsymbol \Phi^{\mathrm{T}} \mathbf{D}^{-1} \boldsymbol \Phi = \Phi^{\mathrm{T}} \Phi \otimes \text{diag}(\tau_1^{-2},\dots,\tau_p^{-2})$, which will be sparse for any choice of basis functions. In general, solving systems with $\mathbf{Q} + \boldsymbol \Phi^{\mathrm{T}} \mathbf{D}^{-1} \boldsymbol \Phi$ does not have an exploitable structure since $\mathbf{Q}$ is block diagonal yet $\boldsymbol \Phi^{\mathrm{T}} \mathbf{D}^{-1} \boldsymbol \Phi$ is a Kronecker product. However, using orthogonal basis functions (i.e.,\ $\Phi^{\mathrm{T}} \Phi = I_L$) means that $\boldsymbol \Phi^{\mathrm{T}} \mathbf{D}^{-1} \boldsymbol \Phi$ is diagonal, so the first term in (\ref{linearization}) is block diagonal and can be easily inverted in $\mathcal{O}(L p^3)$. 

The linearization also involves $\boldsymbol \Phi^{\mathrm{T}} \mathbf{D}^{-1} \mathbf{S} \mathbf{D}^{-1} \boldsymbol \Phi =(\boldsymbol \Phi^{\mathrm{T}} \mathbf D^{-1} \mathbf{Y}_{\text{data}} )(\mathbf{Y}_{\text{data}}^{\mathrm{T}} \mathbf D^{-1} \boldsymbol \Phi)/m$, where $\mathbf{Y}_{\text{data}}$ is the data matrix with columns of realizations $\mathbf{Y}_1,\dots,\mathbf{Y}_m$.
Using $\text{vec}(ABC) = (C^{\mathrm{T}} \otimes A ) \text{vec}(B)$ again,
\begin{equation}  \label{PhiDinvY}
\boldsymbol \Phi^{\mathrm{T}} \mathbf D^{-1} \mathbf{Y}_i = \text{vec}(\text{diag}(\tau_1^{-2},\dots,\tau_p^{-2})  \text{mat}(\mathbf{Y}_i) \Phi)
\end{equation}
where $\text{mat}(\mathbf{Y}_i)$ is the $p \times n$ matrix with $\text{vec}(\text{mat}(\mathbf{Y}_i))= \mathbf{Y}_i$. So $\boldsymbol \Phi^{\mathrm{T}} \mathbf D^{-1} \mathbf{Y}_{\text{data}}$ can be calculated in $\mathcal{O}(m n pL )$ and has low storage cost $\mathcal{O}(m p L)$. Once $\boldsymbol \Phi^{\mathrm{T}} \mathbf D^{-1} \mathbf{Y}_{\text{data}}$ and the block diagonals of $(\mathbf{Q} + \boldsymbol \Phi^{\mathrm{T}} \mathbf{D}^{-1} \boldsymbol \Phi)^{-1}$ are computed, the block diagonals of the second term in (\ref{linearization}) can be computed in $\mathcal{O}(L p^2 m)$.

To summarize, for an orthogonal basis, only $\{ \mathbf{Q}_1,\dots,\mathbf{Q}_L\},$ $\{\tau_1^{-2},\dots,\tau_p^{-2}\}$, and $\boldsymbol \Phi^{\mathrm{T}} \mathbf D^{-1} \mathbf{Y}_{\text{data}}$  must be stored in memory, and we can compute the block diagonals of (\ref{linearization}) in $\mathcal{O}(Lp^3 + L p^2 m)$.
These $L$ $p \times p$ block matrices are then sent into the FMGL as an array of $L$ ``sample covariance'' matrices, and the solution to the FMGL problem is the next guess for $\{ \mathbf{Q}_1,\dots,\mathbf{Q}_L\}$, and the entire procedure is repeated until $\| \mathbf{Q}^{(j+1)} - \mathbf{Q}^{(j)} \|_{\text{F}} / \|\mathbf{Q}^{(j)}\|_{\text{F}}< \epsilon$, which we set as $\epsilon = 0.05$.

\subsubsection{Initial Guess (Unfused Estimate)}  \label{initguess}

When considering the fusion penalty, a natural initial guess is the corresponding unfused estimate. With $\rho=0$, the main optimization (\ref{blockdiagpenaltyproblem}) amounts to solving the graphical lasso independently by level:
\begin{equation}\label{BGLblockdiag} 
 \argmin_{\mathbf{Q}_{\ell} \succeq 0, \ \ell=1,\dots,L}   \left(   \sum_{\ell=1}^L - \log \det \mathbf{Q}_\ell +\text{tr} 
  \left(  \boldsymbol \Psi_{\ell} \mathbf Q_\ell \right)   + \lambda  \sum_{i \neq j} \left\vert(\mathbf{Q}_{\ell})_{ij} \right\vert \right).
  \end{equation}
The entire algorithm has linear complexity and storage in $L$ in this case.

\subsection{Maximum Likelihood Estimate}  \label{MLE}

It is also easy to obtain the unpenalized maximum likelihood estimates for $\mathbf{Q}_1,\dots,\mathbf{Q}_L$. Recall that $S^{-1} = \argmin \limits_{Q \succeq 0} \ - \log \det Q + \text{tr}(SQ)$, assuming $S$ is nonsingular. This means that our DC algorithm would simply invert the linearization matrix (or each block diagonal of the linearization matrix in the multivariate case) rather than substitute it into the graphical lasso. Again, $\mathbf{Q}_1,\dots,\mathbf{Q}_L$ are independent, which implies the same storage and complexity as the unfused estimate in Section \ref{initguess}. This algorithm takes many more DC iterations to converge than either of the regularized estimates.

\subsection{Estimating the Error Variances}  \label{errorvariance_estimate}

Given a single variable with sample covariance $S$, we minimize the following function jointly over $\tau^2$ and a few parameters describing a diagonal matrix $Q$:
\[
\log \det \left(Q + \tau^{-2} \Phi^{\mathrm{T}} \Phi  \right) 
  - \log \det Q  -  \text{tr} \left( \tau^{-4} \Phi^{\mathrm{T}} S \Phi 
  \left( Q + \tau^{-2} \Phi^{\mathrm{T}} \Phi \right) ^{-1}  \right) 
  + n \log \tau^2 +  \tau^{-2} \text{tr}(S).
\]
Note that this expression can be simplified with orthogonal bases, and the trace terms can be quickly computed as squared Frobenius norms. The diagonal parameterization of  $Q$ and parameter estimation is further discussed in the supplementary material. We record values for $\tau_1^2,\dots,\tau_p^2$ in Table 1 in the supplementary material which are used throughout the rest of the paper. 
A similar approach used in \citet{BGL} was found to be successful in recovering error variances even with a misspecified spatial covariance structure.

\subsection{Cross-Validation}  \label{cvpenalty_estimate}

First we describe the cross-validation procedure for a pair of penalty parameters $(\lambda,\rho)$.
Suppose we use $k$ folds and consider $t$ arbitrary pairs of penalties represented by $\{\Lambda_1,\dots,\Lambda_t \}$. Let $\hat Q_{\Lambda_j}(\mathbf{S})$ be the estimate we get from applying our algorithm with empirical covariance $\mathbf{S} = \frac{1}{m} \sum_{i=1}^m \mathbf{Y}_i\mathbf{Y}_i^{\mathrm{T}}$ and penalty pair $\Lambda_j$. For $A \subseteq \{1,\dots,m\}$, let $\mathbf{S}_A = |A|^{-1} \sum_{i \in A} \mathbf{Y}_i\mathbf{Y}_i^{\mathrm{T}}$. We seek $\Lambda$ so that $\alpha(\Lambda)= \ell(\hat{\mathbf{Q}}_\Lambda(\mathbf{S}), \mathbf{S})$ is small, where 
\begin{equation} \label{finallikelihood}
  \ell(\mathbf{Q},\mathbf{S}) =\log \det \left( \mathbf{Q} + \boldsymbol \Phi^{\mathrm{T}} \mathbf D^{-1} \boldsymbol \Phi  \right) - \log \det \mathbf{Q}
- \text{tr} \left(  \boldsymbol \Phi^{\mathrm{T}} \mathbf D^{-1} \mathbf{S} \mathbf D^{-1} \boldsymbol \Phi  ( \mathbf{Q} + \boldsymbol \Phi^{\mathrm{T}} \mathbf D^{-1} \boldsymbol \Phi)^{-1}  \right)
\end{equation}
is the unpenalized likelihood function in (\ref{Qlikelihood.pl}). The cross-validation approach is to partition $\{1,\dots,m\}$ into disjoint sets $\{A_1,\dots,A_k\}$ and select $\hat \Lambda = \argmin \limits_{ \Lambda \in \{\Lambda_1,\dots,\Lambda_t \}} \hat \alpha(\Lambda)$ where \\ $\hat \alpha(\Lambda) = k^{-1} \sum_{i=1}^k \ell(\hat{\mathbf{Q}}_\Lambda (\mathbf{S}_{A_i^c}),\mathbf{S}_{A_i} )$.

Jointly searching over $\lambda$ and $\rho$ can quickly become unwieldy even when considering a small combination of sparsity and fusion penalties, as noted by \cite{JGL}, who instead suggest a dense search for $\lambda$ with $\rho=0$ fixed and then a search for $\rho$ with that sparsity value fixed. 
The individual cross-validation for either $\lambda$ or $\rho$ follows the same idea: whichever penalty parameter has the lowest negative log-likelihood average across folds is selected.

\section{Data Analysis}

%[xxx: Dorit, need your help here! Good place to introduce the science problem and data specifics.]

This section is broken into two main, but related, application and validation efforts. 
Both are done with the lens of the climate data problem: we begin with some exploratory analyses of the climate dataset, deriving reasonable basis functions and providing discussion to guide intuition for the ensuing model application. 
In the supplementary material, we detail a simulation study that tests our ability, under a similar setup as the climate example, to recover meaningful and relevant graphs for coefficients at different levels of basis functions under realistic assumptions on possible graph structures. Results from this simulation study are promising and lead us to expect reasonable results with the real climate data as well. 
The final section in the body of this paper provides the full analysis of our model on the Community Atmosphere Model (CAM) data along with scientific interpretations of recovered graphical structures and some implied covariance and cross-covariance patterns. Code which outlines the data analysis procedure is available at \texttt{github.com/mlkrock/MultivariateBasisGraphicalLasso}.

\subsection{Data Description and Exploratory Analyses}  \label{camintro}

We apply our method to a large climatological dataset from an ensemble study conducted at the National Center for Atmospheric Research (NCAR). 
Climate variability is typically assessed by examining a collection of numerical climate model simulations, which are computationally and economically expensive to produce. Relationships between variables at different spatial scales are crucial for scientific investigations; hence a scalable statistical model which can simulate multivariate processes could be a powerful tool for climatologists. Our method allows for efficient emulation and straightforward interpretation of complex geophysical model variable relationships potentially filling this niche.

A climate model ensemble is typically a collection of climate simulations from the same numerical model using various initial conditions; ours is an extended version of the ensemble described in \cite{gmd-8-2829-2015} with $m=343$ members.
Data are recorded at $n=$ \numprint{48,602} gridded spatial locations over the globe. 
There are a total of 164 variables available, the majority of which are three-dimensional, meaning they have a third dimension corresponding to 30 vertical atmospheric levels. 
For our study we only consider the two-dimensional surface variables, and a subset thereof. 
First, the variables are on different scales, so they are standardized with a pixelwise empirical mean and pixelwise empirical standard deviation. 
Histograms and Q-Q plots were consulted to remove strongly non-normal variables. 
Note that our data are yearly-averaged quantities, so a Gaussian assumption is generally reasonable. 
Potential variables were also removed if they were very strongly correlated, suggesting redundant information (e.g., when vectorized across space and realizations, the absolute correlation between two processes was above $0.9$). 
We settled upon the $p=40$ variables listed in Table \ref{tab:vardescription}, which are grouped into five categories: aerosol variables, cloud variables, flux variables, precipitation variables, and transport/state variables. Throughout the rest of the document, aerosol variables are colored red, cloud variables are colored blue, flux variables are colored green, precipitation variables are colored purple, and transport/state variables are colored grey.

\begin{table}[h!]
\centering
\footnotesize
\begin{tabular}{|c|c|c|c| }  \hline
Variable & Description & Units & Category \\ \hline
AODVIS & Aerosol optical depth  & 550 nm & Aerosol  \\
BURDEN1 & Aerosol burden mode 1 & kg/m$^2$ & Aerosol \\
BURDEN2 & Aerosol burden mode 2 & kg/m$^2$ & Aerosol \\
BURDEN3 & Aerosol burden mode 3 & kg/m$^2$ & Aerosol \\
BURDENBC & Black carbon aerosol burden & kg/m$^2$ & Aerosol \\
BURDENPOM & POM aerosol burden & kg/m$^2$ & Aerosol \\
BURDENSEASALT & Seasalt aerosol burden & kg/m$^2$  & Aerosol \\
BURDENSO4 & Sulfate aerosol burden & kg/m$^2$ & Aerosol \\
BURDENSOA & SOA aerosol burden & kg/m$^2$ & Aerosol \\
CDNUMC & Vertically-integrated droplet concentration & 1/m$^2$ & Cloud \\
CLDHGH & Vertically-integrated high cloud & fraction & Cloud\\
CLDMED & Vertically-integrated mid-level cloud & fraction & Cloud\\
CLDTOT & Vertically-integrated total cloud & fraction & Cloud \\
FLDS & Downwelling longwave flux at surface & W/m$^2$ & Flux \\
FLNS & Net longwave flux at surface & W/m$^2$ & Flux \\
FLNSC & Clearsky net longwave flux at surface & W/m$^2$ & Flux \\
FLNT & Net longwave flux at top of model & W/m$^2$ & Flux \\
FLNTC & Clearsky net longwave flux at top of model & W/m$^2$ & Flux \\
FSDS & Downwelling solar flux at surface & W/m$^2$ & Flux \\
FSDSC & Clearsky downwelling solar flux at surface & W/m$^2$ & Flux \\
FSNS & Net solar flux at surface & W/m$^2$ & Flux \\
FSNSC & Clearsky net solar flux at surface & W/m$^2$ & Flux \\
FSNTC & Clearsky net solar flux at top of model & W/m$^2$ & Flux \\
FSNTOA & Net solar flux at top of atmosphere & W/m$^2$ & Flux \\
LHFLX & Surface latent heat flux & W/m$^2$  & Flux \\
LWCF & Longwave cloud forcing & W/m$^2$  & Cloud \\
PBLH & PBL height & W/m$^2$ & Transport/State \\ 
PS & Surface pressure & Pa & Transport/State \\
QREFHT & Reference height humidity & kg/kg & Precipitation \\
SHFLX & Surface sensible heat flux & W/m$^2$ & Flux \\
SWCF & Shortwave cloud forcing & W/m$^2$ & Cloud \\
TAUX & Zonal surface stress & N/m$^2$ & Transport/State \\
TAUY & Meridional surface stress & N/m$^2$ & Transport/State \\
TGCLDCWP & Total grid-box cloud water path (liquid and ice) & kg/m$^2$ & Cloud \\
TGCLDIWP & Total grid-box cloud ice water path & kg/m$^2$ & Cloud \\
TGCLDLWP & Total grid-box cloud liquid water path & kg/m$^2$ & Cloud  \\
TMQ & Total vertically integrated precipitable water & kg/m$^2$ & Precipitation \\
TREFHT & Surface air temperature at reference height  & K & Transport/State  \\
U10 & 10m wind speed & m/s & Transport/State \\
PRECT & PRECL Large-scale (stable) precipitation rate (liq + ice)  & m/s & Precipitation  \\
   &  plus PRECC Convective precipitation rate (liq + ice)  & & \\ \hline
\end{tabular}
\caption{Variable descriptions. Note that analysis happens on standardized, unitless data.}% See \url{https://www.cesm.ucar.edu/models/cesm1.2/cam/docs/ug5_3/hist_flds_fv_cam5.html} for more description.}
\label{tab:vardescription}
\end{table}

With the $p=40$ variables in hand, our approach relies on first specifying a set of spatial basis functions. 
We construct such functions as empirical orthogonal functions (EOFs) \citep{wikle2010}, which are widely used in the atmospheric and climate sciences. 
Typically, EOFs are used in a temporal context with a single variable. 
Let's consider a single spatiotemporal variable and suppose we have a matrix $B$ of data with rows indexing $n$ spatial locations and columns indexing $t$ time points. 
If $B = U D V^{\mathrm{T}}$ is the (economy) SVD of the data matrix, the columns of the orthogonal matrix $U$, referred to as EOFs, represent the normalized eigenvectors of the process empirical covariance matrix $B B^{\mathrm{T}}$. 
%Note that the percent of variance explained by the first $k$ EOFs is  given by $(\sum_{i=1}^k D_{ii}^2)/(\sum_{i=1}^{t} D_{ii}^2)$.

\begin{figure}[t]
  \begin{subfigure}[b]{0.32\linewidth}
    \centering
    \includegraphics[width=\linewidth]{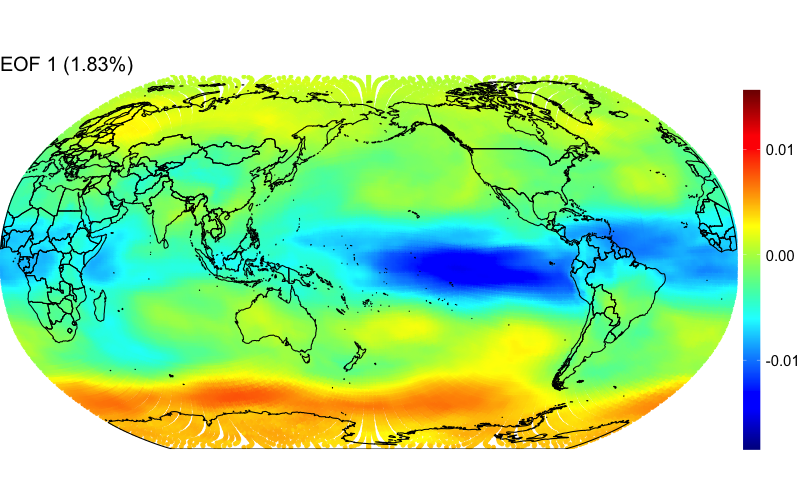}
      \end{subfigure} 
        \begin{subfigure}[b]{0.32\linewidth}
    \centering
    \includegraphics[width=\linewidth]{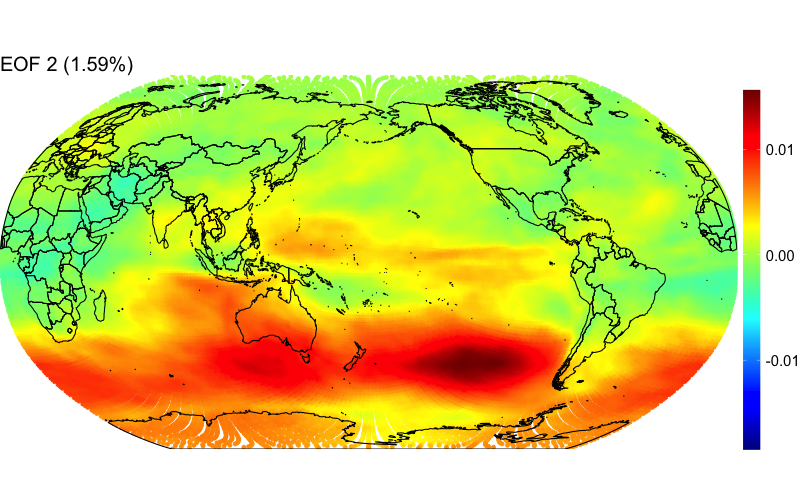}
      \end{subfigure} 
      \begin{subfigure}[b]{0.32\linewidth}
      \centering
    \raisebox{-.15\height}{\includegraphics[width=\linewidth]{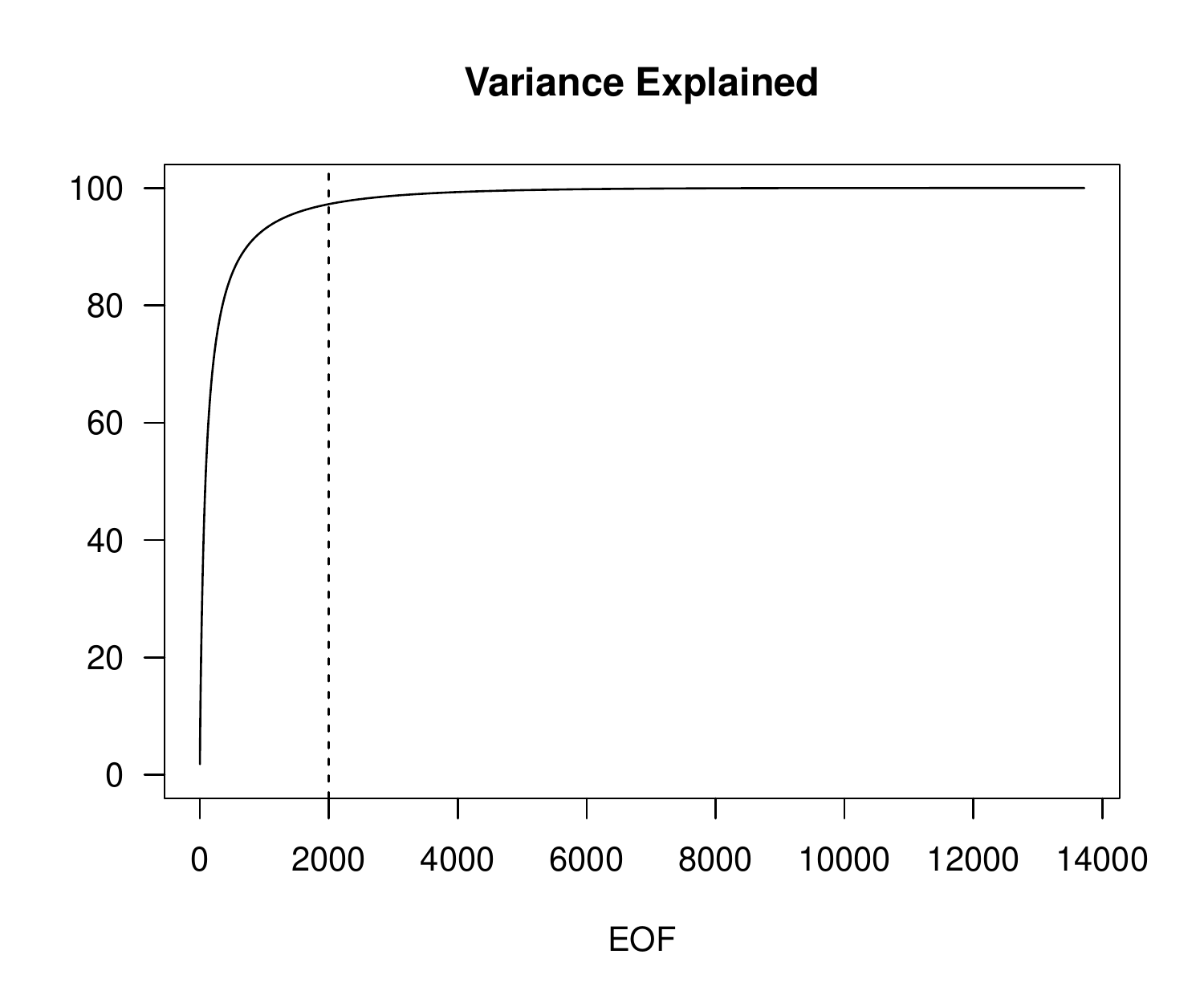}}
          \end{subfigure} 
  \caption{First two pooled EOFs of the standardized CAM data. The two EOFs account for 1.83\% and 1.59\% of the total variability of all 40 variables, respectively. Rightmost plot shows the cumulative percentage of variability explained by the EOFs, with $L=2000$ EOFs capturing 97.2\% of the total variance.}
  \label{EOFpix}
\end{figure}

We compute the (economy) SVD of the $n \times pm  = \numprint{48,602} \times \numprint{13,720}$ data matrix where a row corresponds to a spatial location and contains the $p$ standardized variables ordered sequentially by realization. 
Such an approach can be thought of as generating {\it pooled} EOFs that describe common structure seen amongst all variables. 
Using the same notation $UDV^{\mathrm{T}}$ for the SVD, we follow common practice and take the first $L$ columns of $U$ to form our $n \times L$ basis matrix $\Phi$. 
%Figure \ref{EOFvarianceexplained} shows that truncating after $L=2000$ EOFs is a reasonable tradeoff between using a relatively small number of basis functions and explaining sufficient variance (97.2\%). 
Exploratory analysis suggests truncating after $L=2000$ EOFs is a reasonable tradeoff between using a relatively small number of basis functions and explaining sufficient variance (97.2\%). 
The first two pooled EOFs are displayed in Figure \ref{EOFpix}. We emphasize that pooling variables together to create EOFs is nontraditional and explains why the first two EOFs account for such little variability.

%\begin{figure}[h!]
%\centering
%\includegraphics[scale=0.22]{figs/EOFvarianceexplained}
%\caption{Percentage variance explained from using various numbers of EOFs. 97.2\% of the variance is explained in the first 2000 EOFs.}
%\label{EOFvarianceexplained}
%\end{figure}

With this formulation it is important to clarify the role of the additive error term. 
Here an interpretation as a traditional spatial statistical nugget effect is not well-motivated and instead we think of $\boldsymbol \varepsilon$ as a fine-scale process which is at smaller scale than the EOFs and absorbs the remaining variability unexplained by the pooling of variables. 
As noted in \cite{wikle2010}, if enough eigenvectors are used to explain sufficient variation, then it is reasonable to assume that the EOF residuals are uncorrelated in space. 
This motivates the white noise assumption on $\boldsymbol \varepsilon$ which in turn suggests independently estimating $\tau_1^2,\dots,\tau_p^2$ using the procedure described in Section \ref{errorvariance_estimate}. Estimated values for $\tau_1^2,\dots,\tau_p^2$ and additional interpretations are shown in the supplementary material.

\subsection{CAM Data Analysis}

We proceed to the data analysis using our estimates for $\tau^2_1,\dots,\tau^2_p$ and the basis setup  from Section \ref{camintro}.
 The remaining question is what penalty parameters to use.
With a cross-validation attempt (see supplementary material) %and the previous section's cross validation results 
and similar difficulties with penalties encountered in \cite{JGL} in mind, we proceed by fixing several penalty pairs and examining the resulting modeling implications. Ideally, we would select a model with a sensible, interpretable graphical neighbor structure over levels. Differences between models with different graph structures may be minor as different graphs can give approximately the same correlation structure.
In the supplementary material and remainder of this document, we display several results for $\lambda=20$ since this looked like an inflection point in a plot of $\lambda$ versus the total graph sparsity percentage (see Figure 6(a) in supplementary material). %The other plot in Figure \ref{qtotalsparsitypct} shows how the sparsity of the graphs corresponding to $\lambda=20$ changes over levels. 
We also will occasionally compare $\lambda=20$ results with those from $\lambda=1$ to give an idea of how the implied graph structure changes with different penalties.

\begin{figure}[h!] 

  \begin{minipage}[b]{0.32\linewidth}
    \centering
    \includegraphics[width=\linewidth]{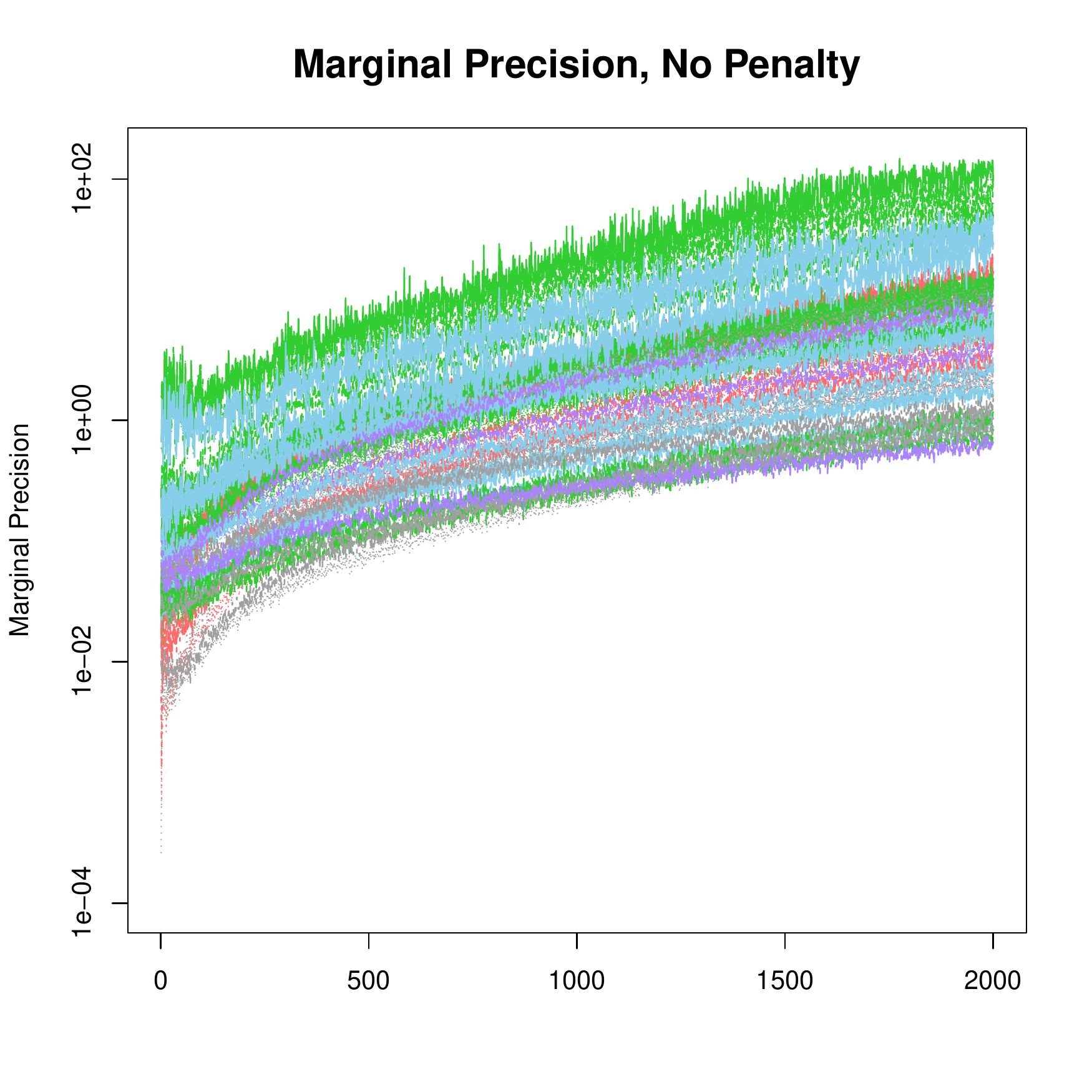} 
  \end{minipage} 
    \begin{minipage}[b]{0.32\linewidth}
    \centering
    \includegraphics[width=\linewidth]{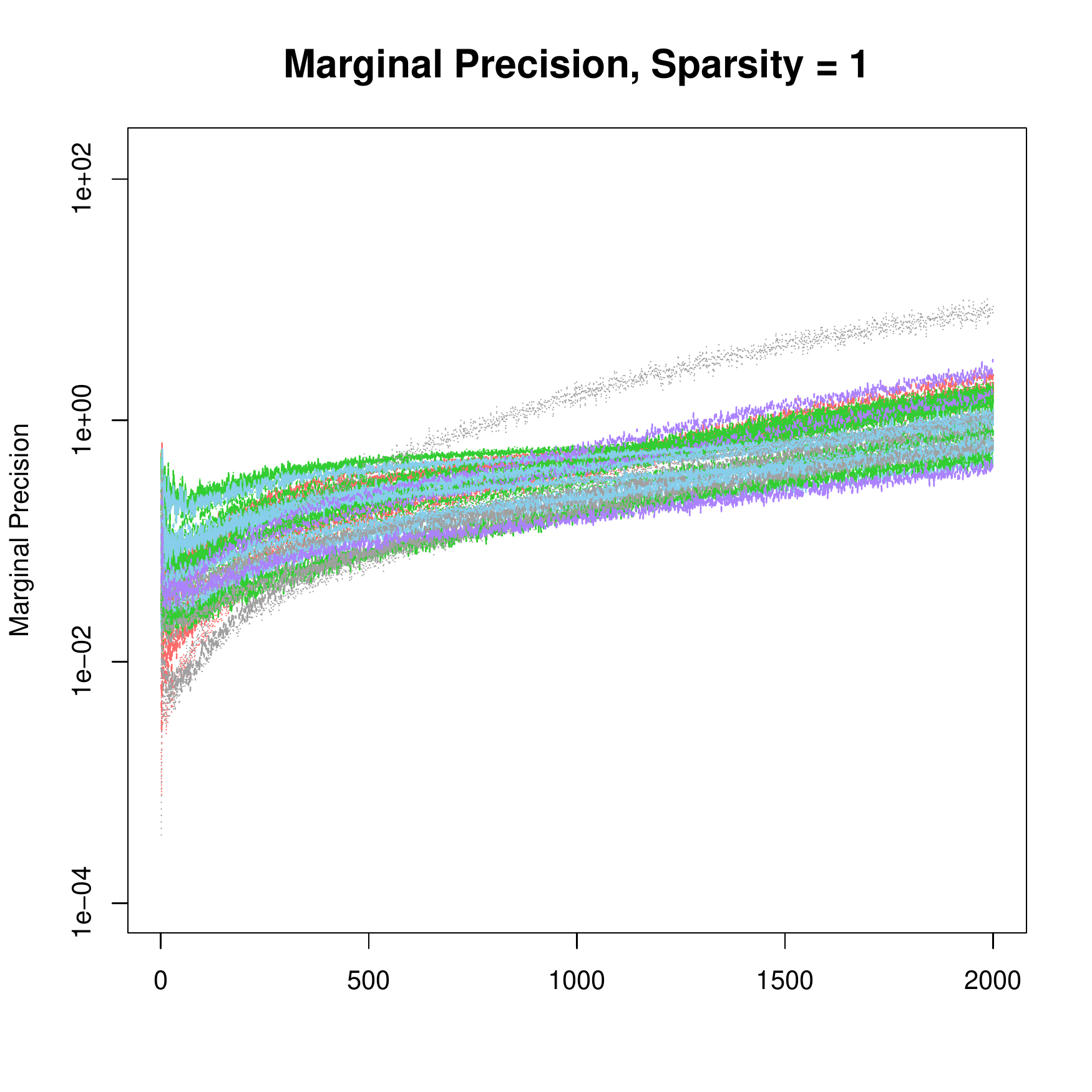} 
  \end{minipage} 
     \begin{minipage}[b]{0.32\linewidth}
    \centering
    \includegraphics[width=\linewidth]{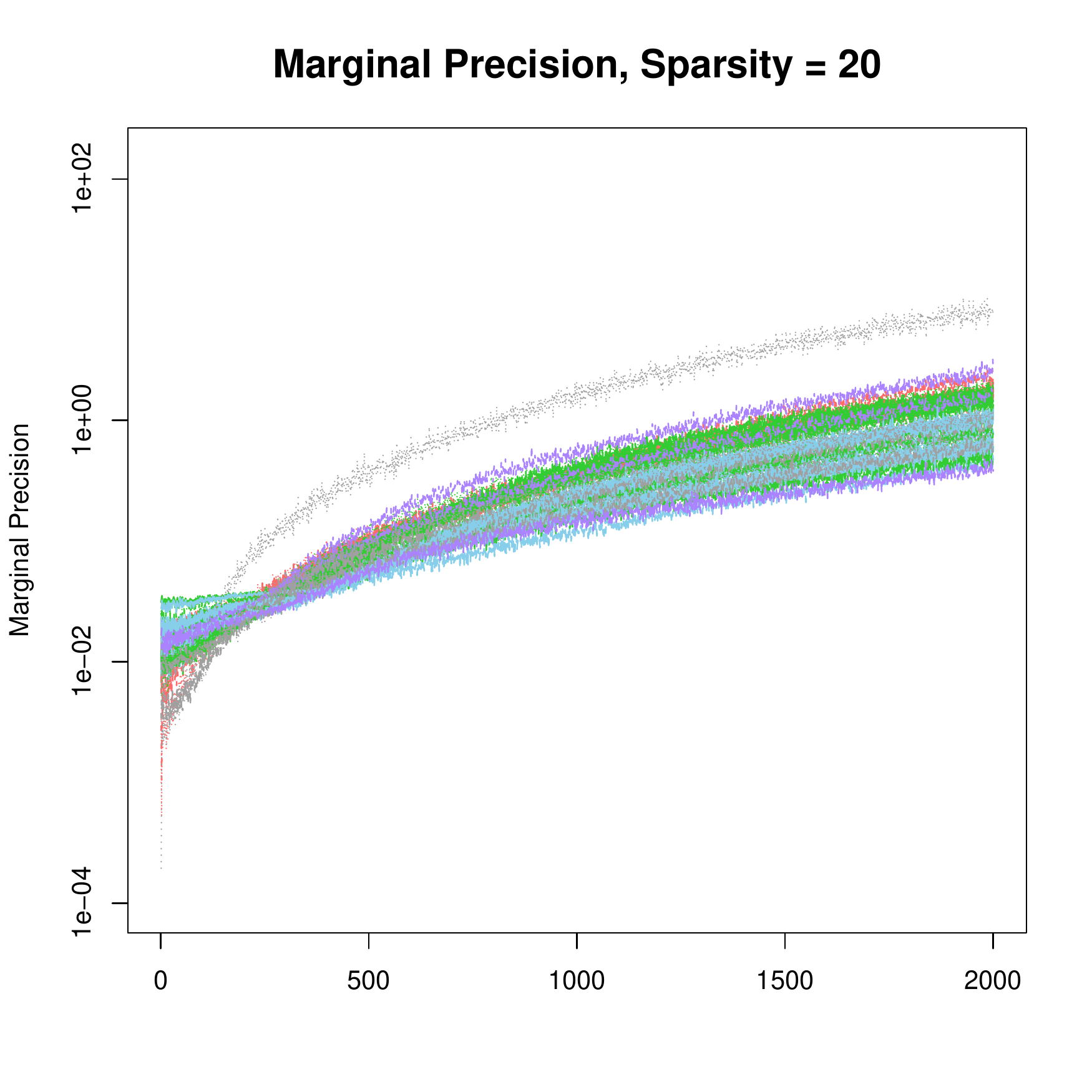} 
  \end{minipage} 
  
  \vspace{-.25in}
  
      \begin{minipage}[b]{0.32\linewidth}
    \centering
    \includegraphics[width=\linewidth]{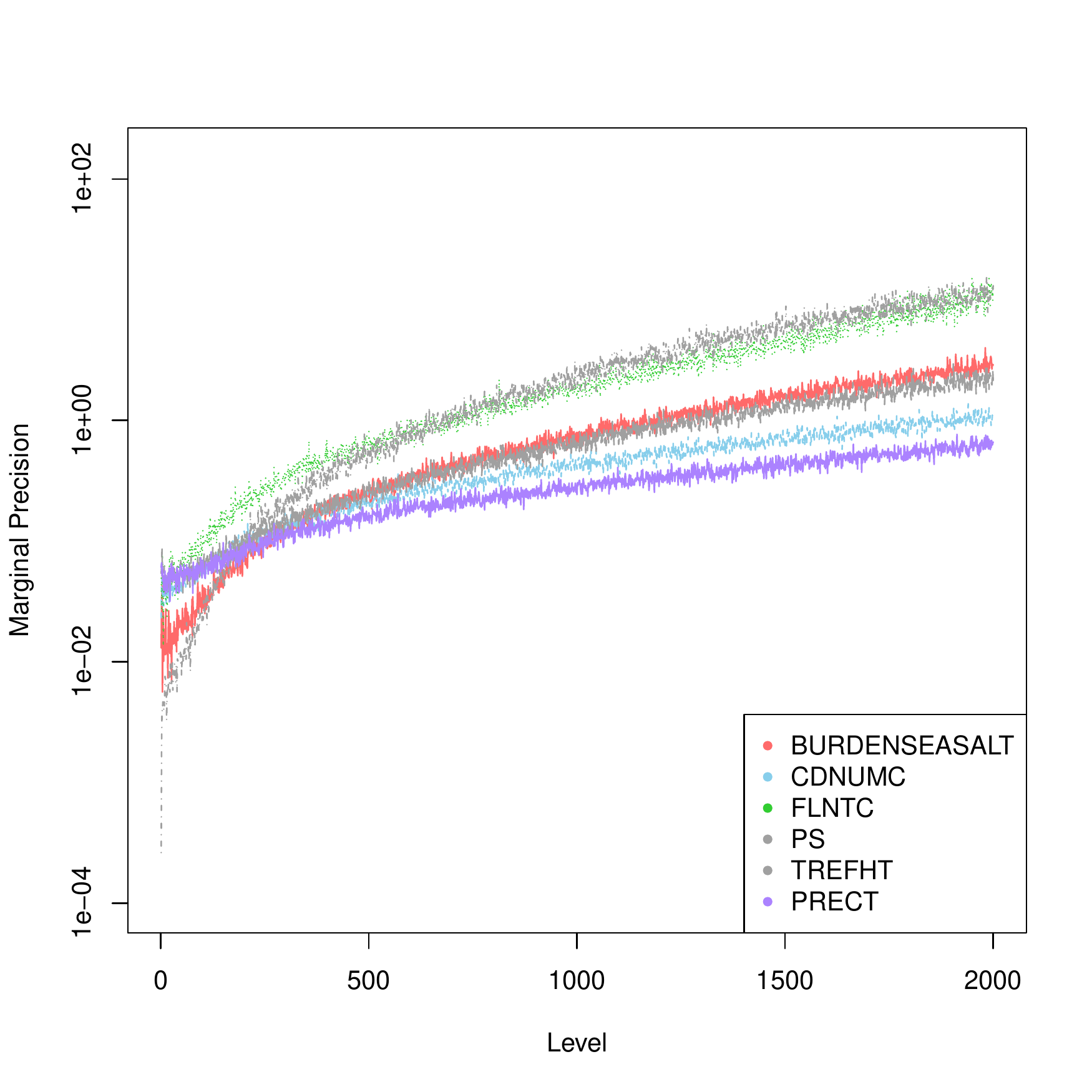} 
  \end{minipage} 
    \begin{minipage}[b]{0.32\linewidth}
    \centering
    \includegraphics[width=\linewidth]{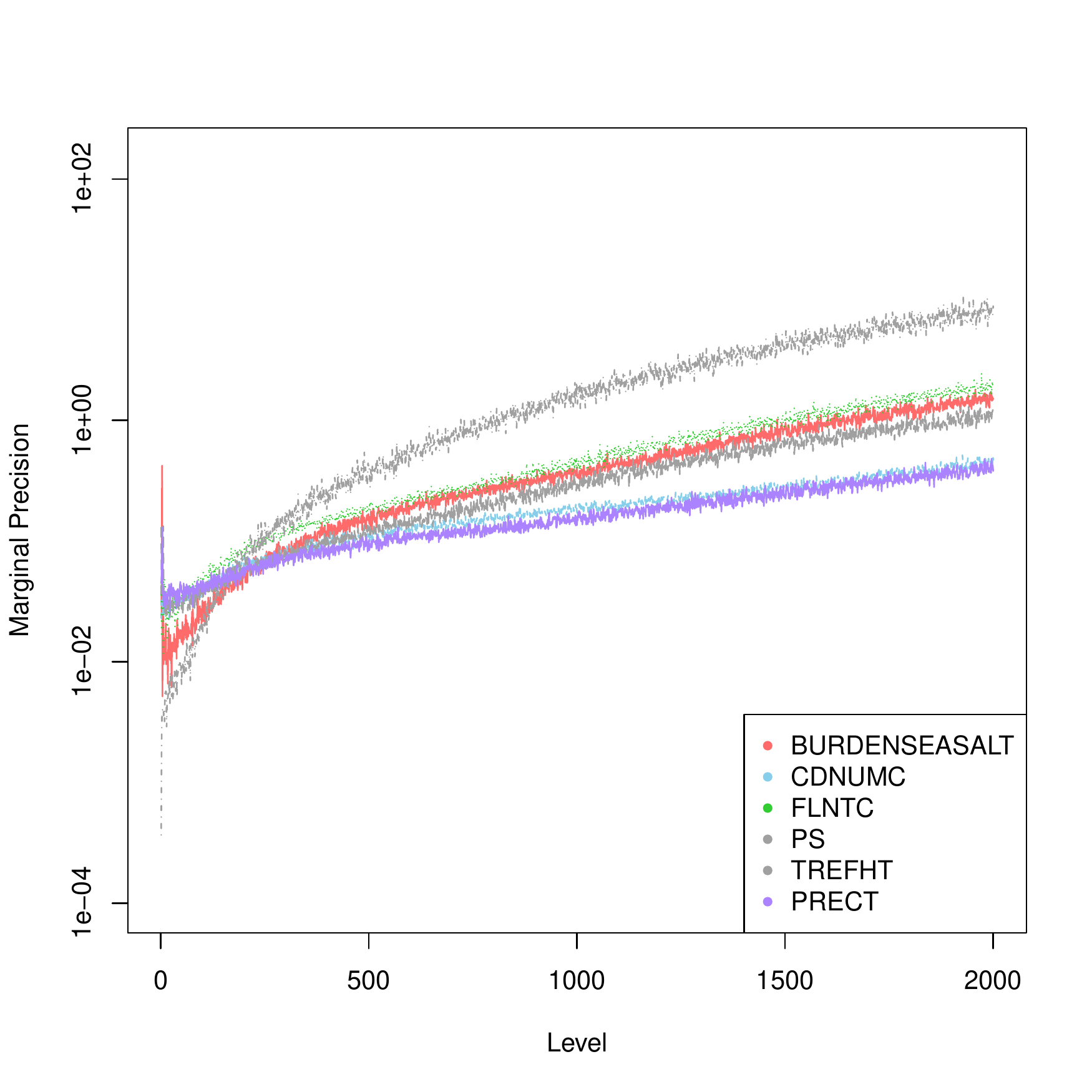} 
  \end{minipage} 
      \begin{minipage}[b]{0.32\linewidth}
    \centering
    \includegraphics[width=\linewidth]{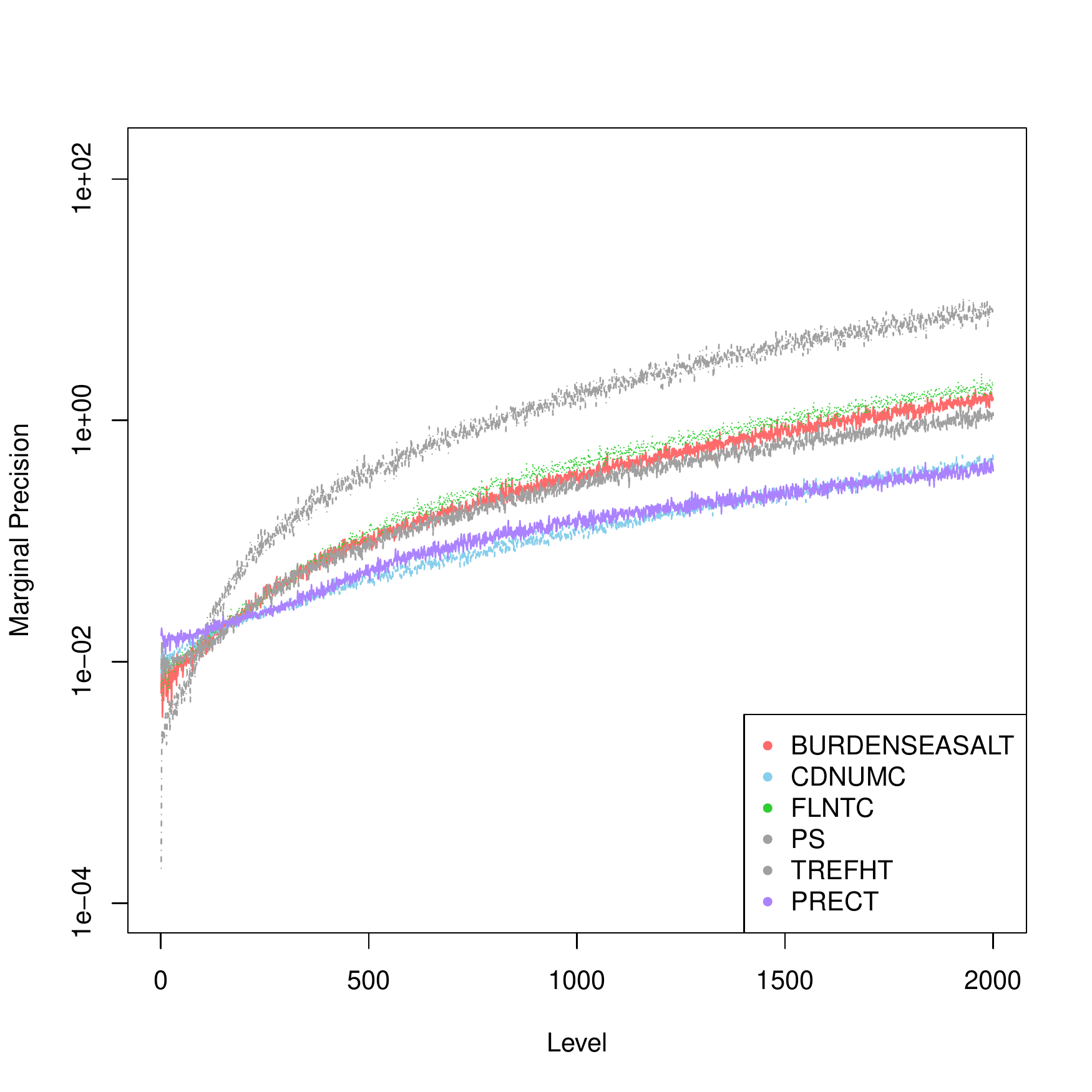} 
  \end{minipage}

  \caption{Marginal precision estimates (i.e.,\ diagonals of $\hat{\mathbf{Q}}_1,\dots,\hat{\mathbf{Q}}_{2000}$) shown by level for various penalty choices. Log scale on $y$-axis. Bottom row shows a subset of six variables from the top row. The grey pressure variable is noticeably smoother than the rest after regularization is added.}
  \label{margprec}
\end{figure}

For the first set of model diagnostics, we examine the estimated precision matrices. Plots of estimated marginal precisions by level are shown in Figure \ref{margprec}. Clearly, the exponential parameterization considered in the simulation study is not well-suited to our data.  Just as in the simulation study, adding a sparsity penalty causes the pressure variable (colored grey) to have the highest marginal precisions at high levels of resolution, as we would expect from the smoothest variable.  Overall, adding a sparsity penalty brings all marginal precisions down by an order of magnitude, with particularly strong shrinkage at lower levels of resolution.  Given the focus on regularization, these results may not seem surprising, but the diagonals of the precision matrix are not penalized in any formulation we have considered. This shrinkage of marginal precisions can be attributed to the larger number of neighbors in the low-penalized graph structures, which means that the (conditional) precision will be higher than in estimates from higher penalties.

\begin{figure}[h!] 

  \begin{minipage}[b]{0.32\linewidth}
    \centering
    \includegraphics[width=\linewidth]{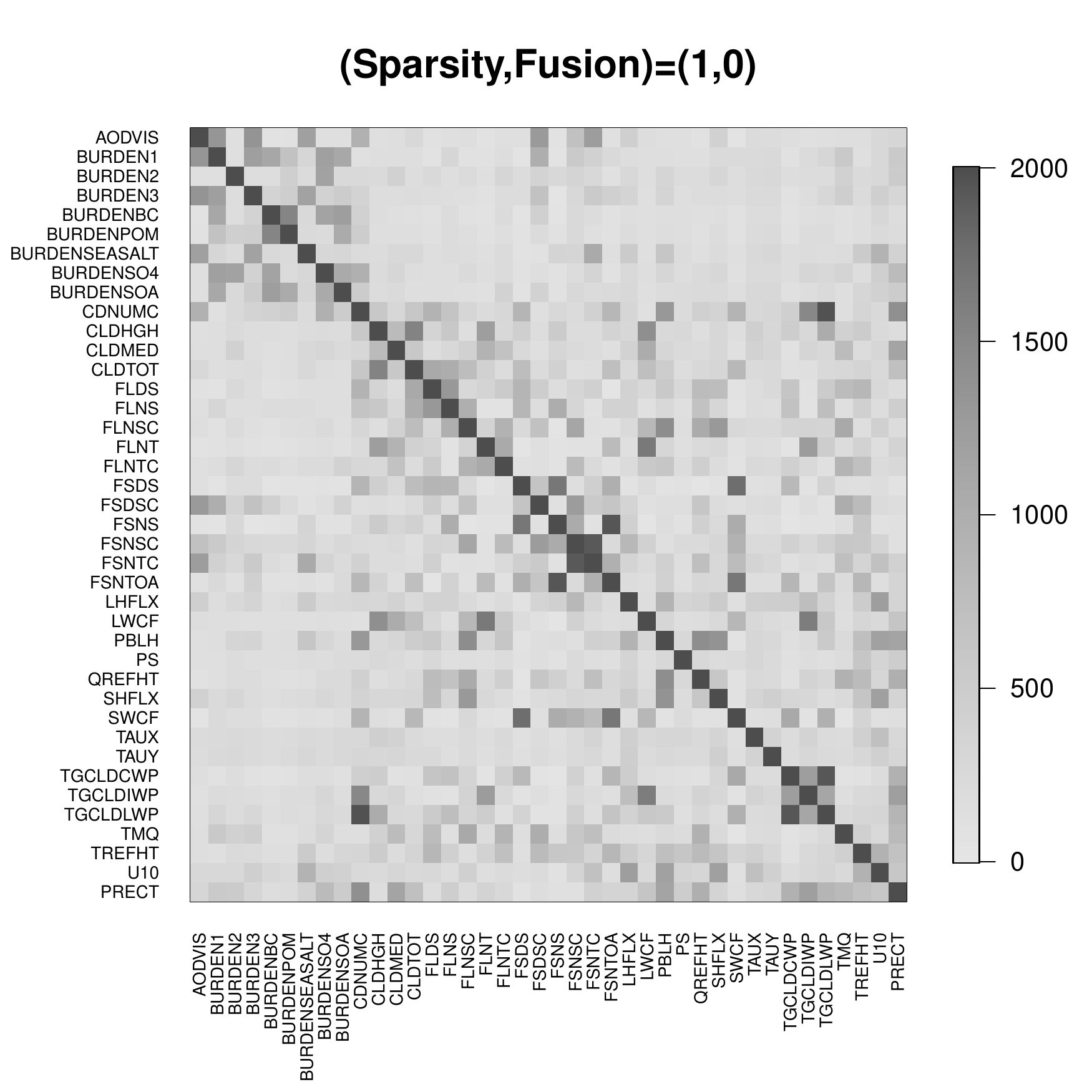} 
  \end{minipage} 
      \begin{minipage}[b]{0.32\linewidth}
    \centering
    \includegraphics[width=\linewidth]{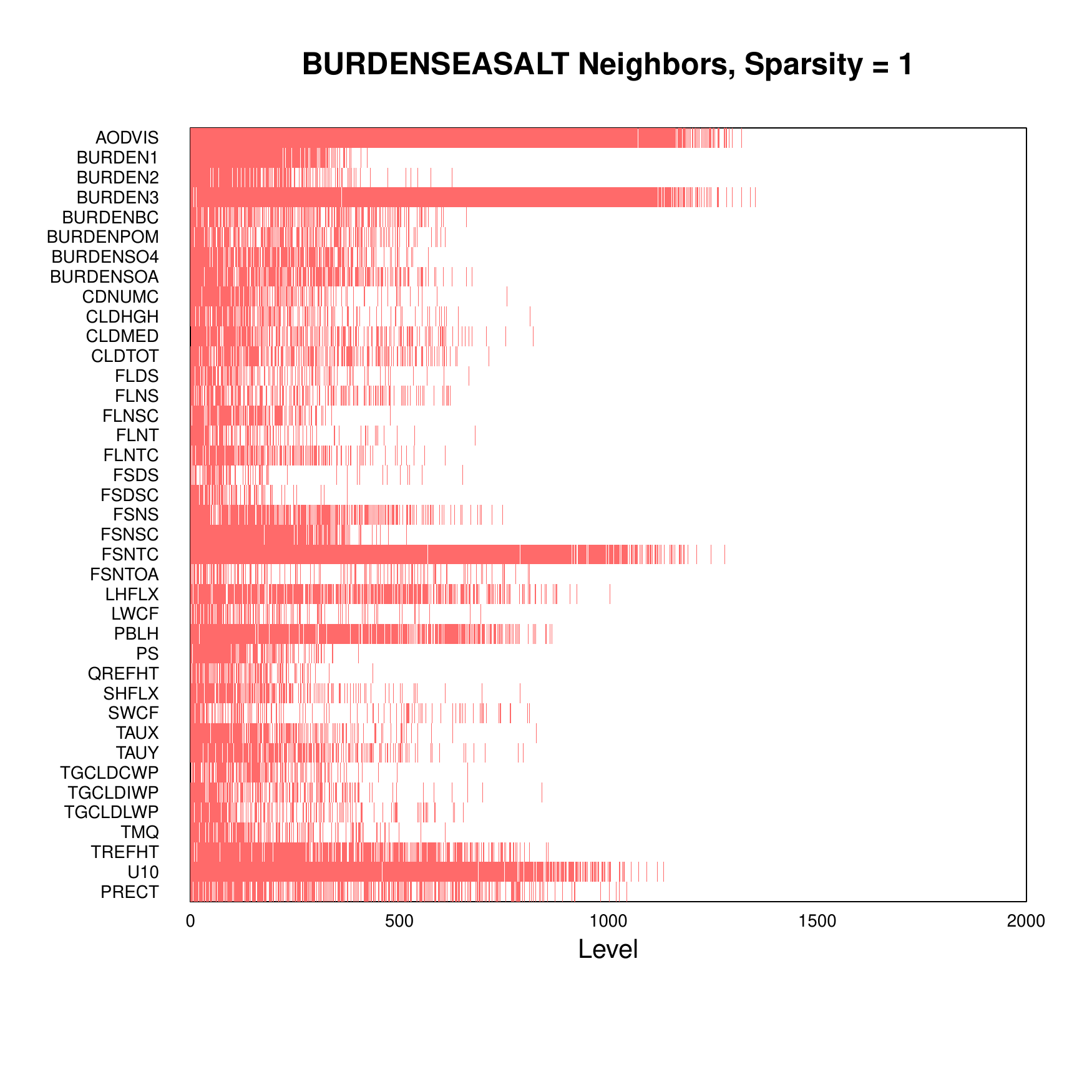} 
  \end{minipage} 
   \begin{minipage}[b]{0.32\linewidth}
    \centering
    \includegraphics[width=\linewidth]{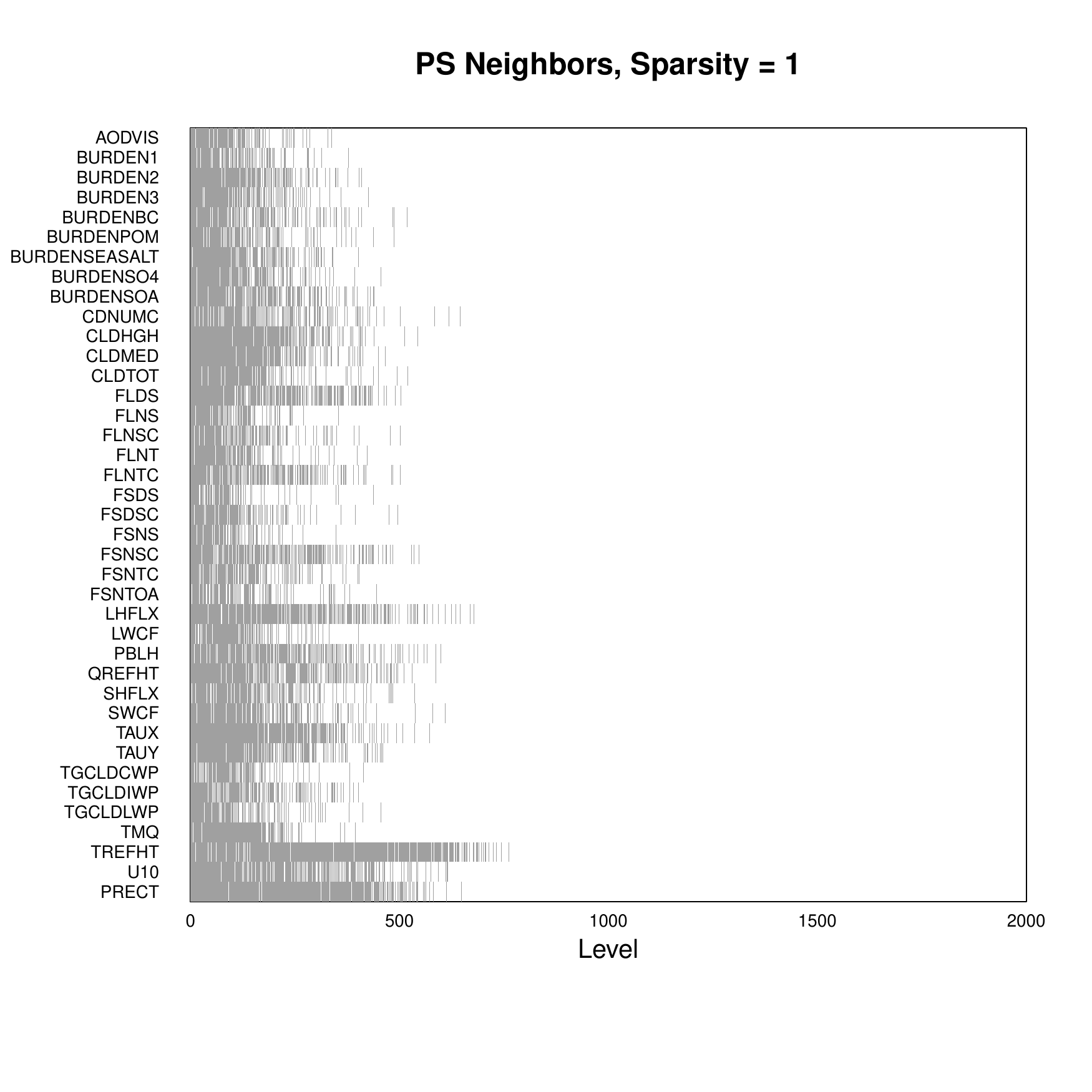} 
  \end{minipage} 
  
    \begin{minipage}[b]{0.32\linewidth}
    \centering
    \includegraphics[width=\linewidth]{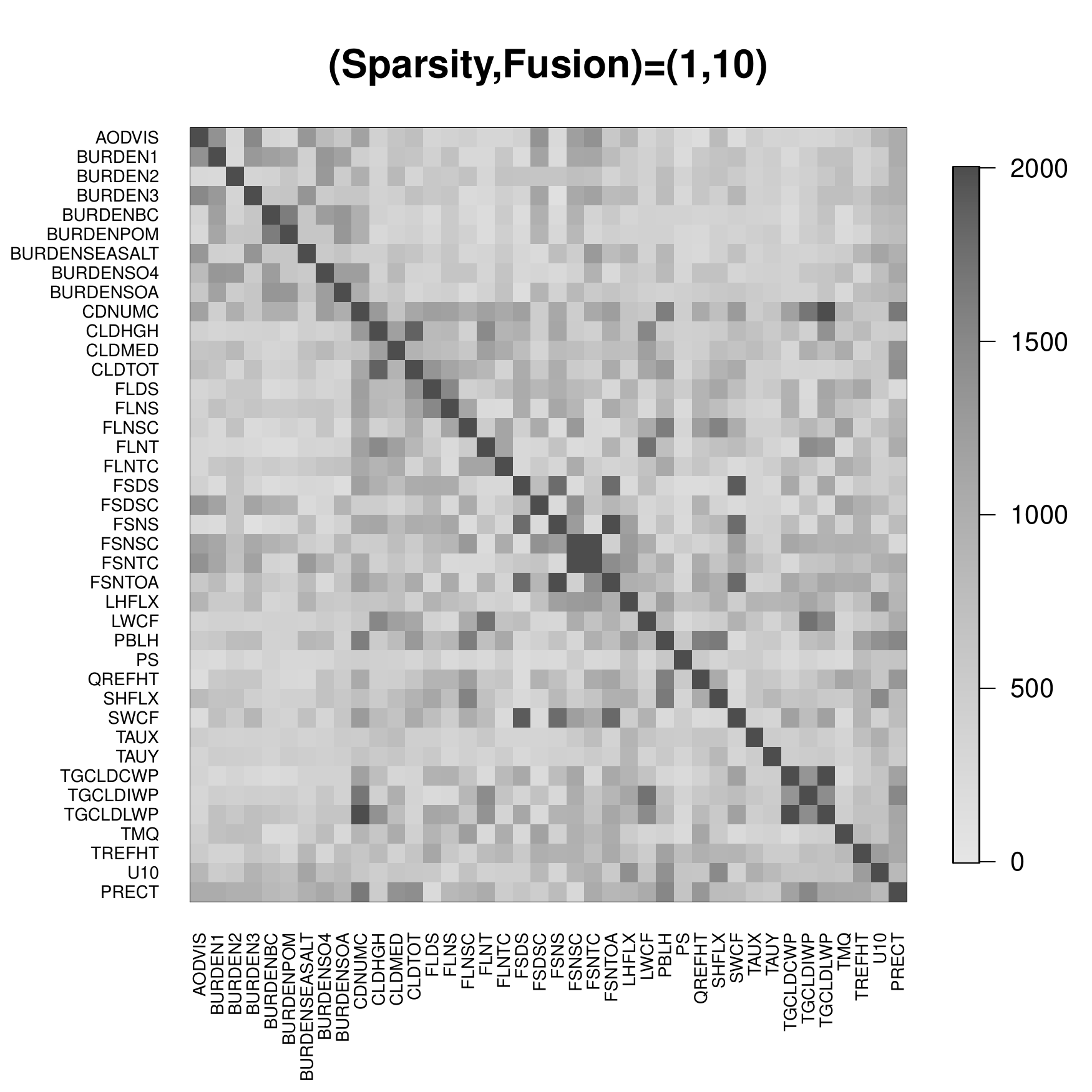} 
  \end{minipage} 
      \begin{minipage}[b]{0.32\linewidth}
    \centering
    \includegraphics[width=\linewidth]{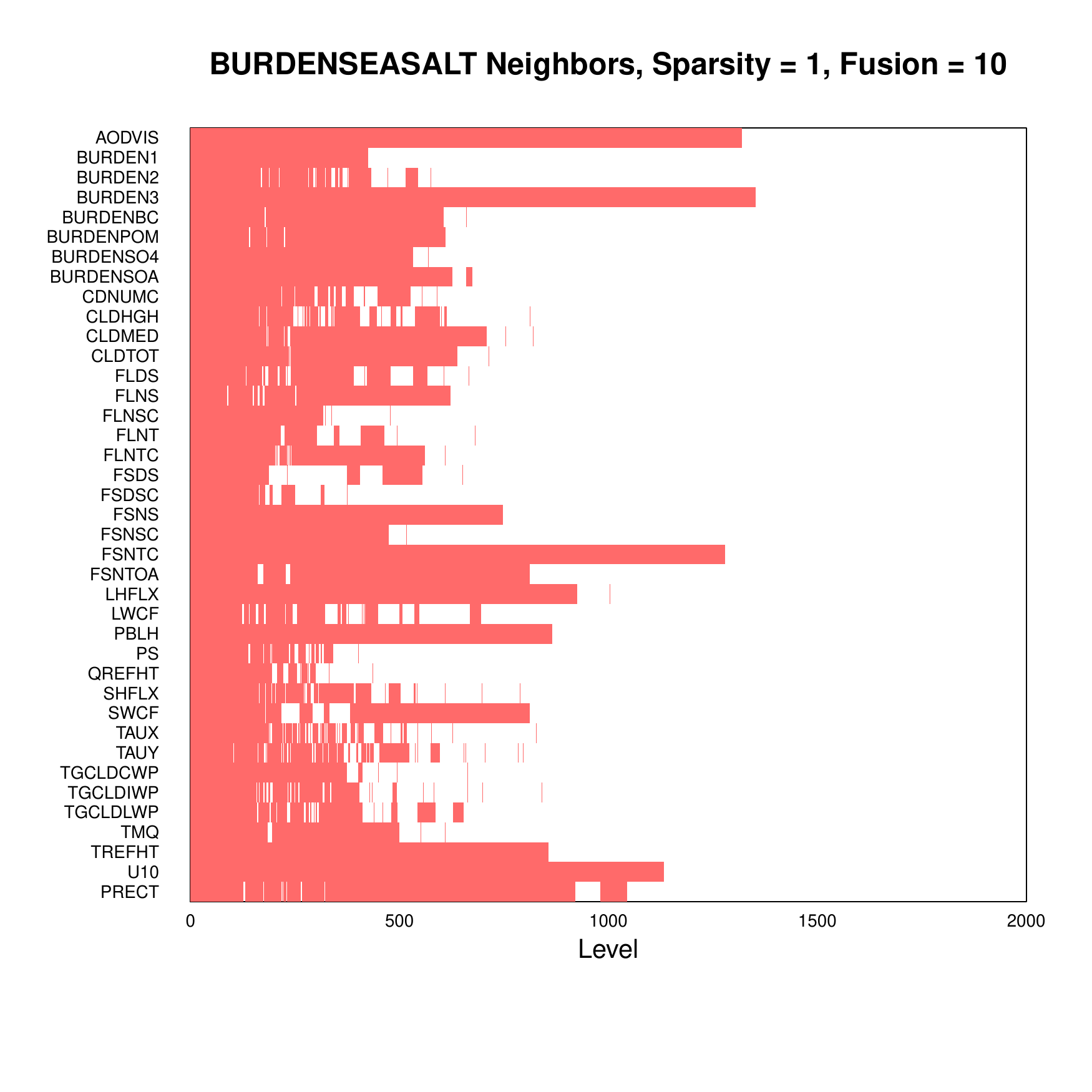} 
  \end{minipage} 
   \begin{minipage}[b]{0.32\linewidth}
    \centering
    \includegraphics[width=\linewidth]{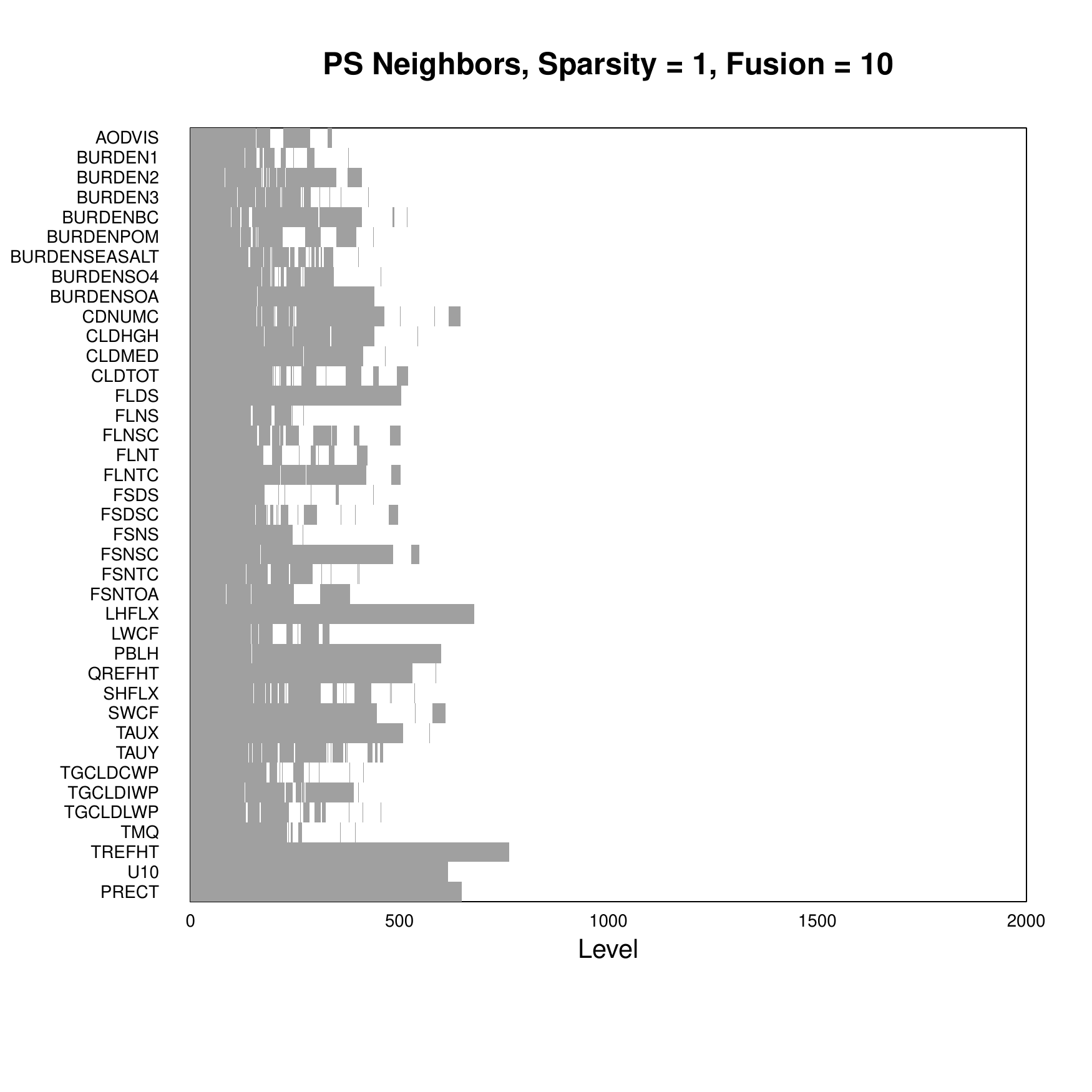} 
  \end{minipage}

  \begin{minipage}[b]{0.32\linewidth}
    \centering
    \includegraphics[width=\linewidth]{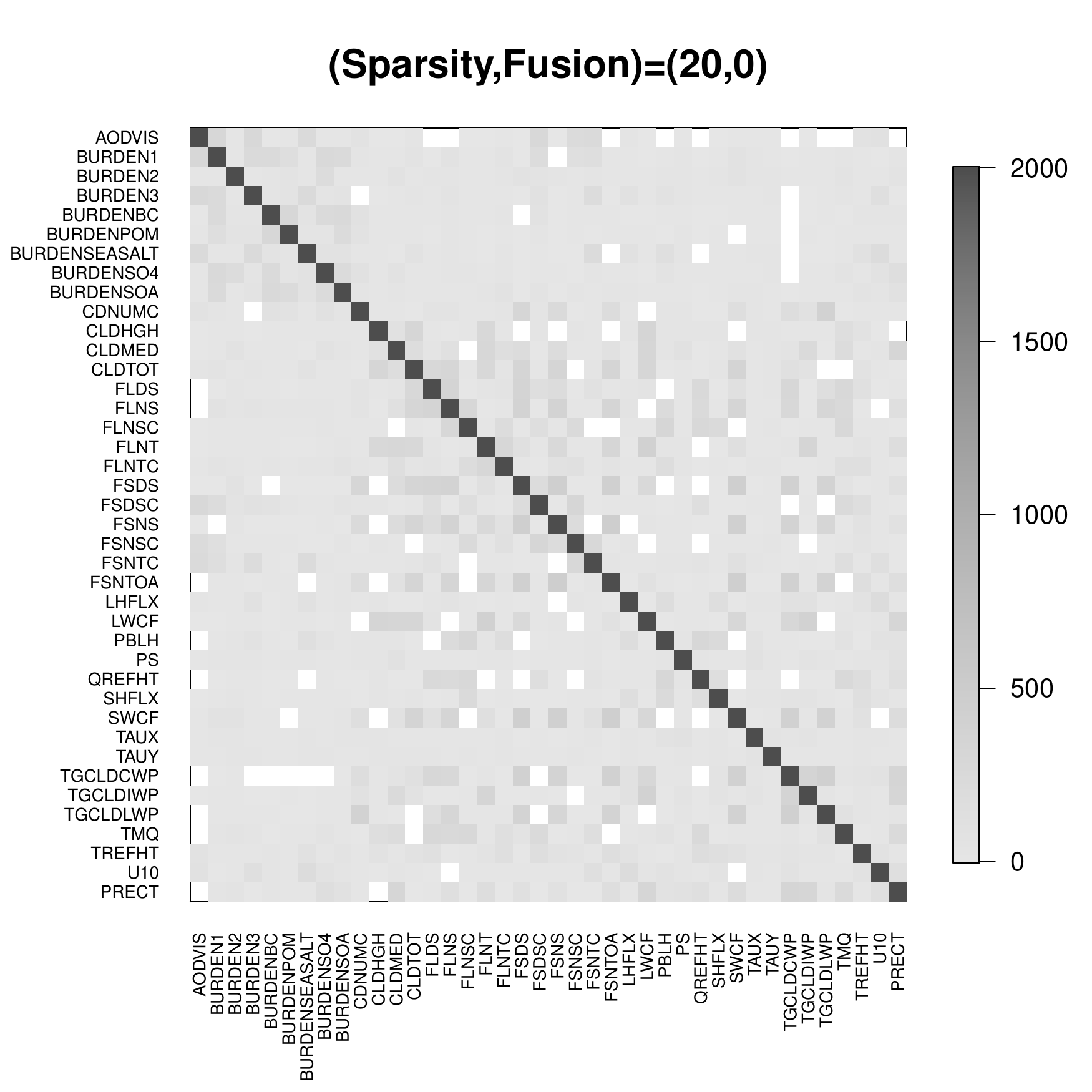} 
  \end{minipage} 
      \begin{minipage}[b]{0.32\linewidth}
    \centering
    \includegraphics[width=\linewidth]{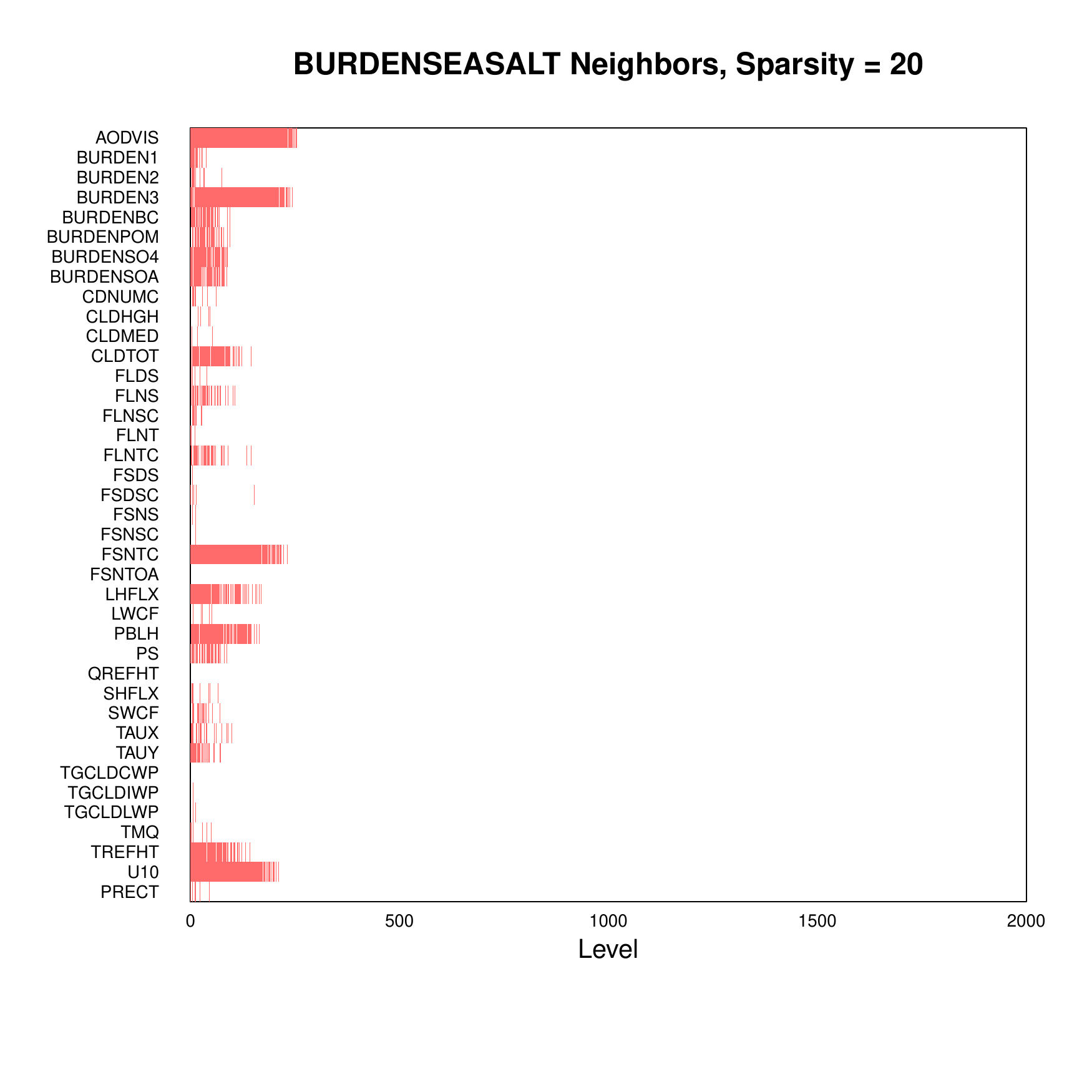} 
  \end{minipage} 
   \begin{minipage}[b]{0.32\linewidth}
    \centering
    \includegraphics[width=\linewidth]{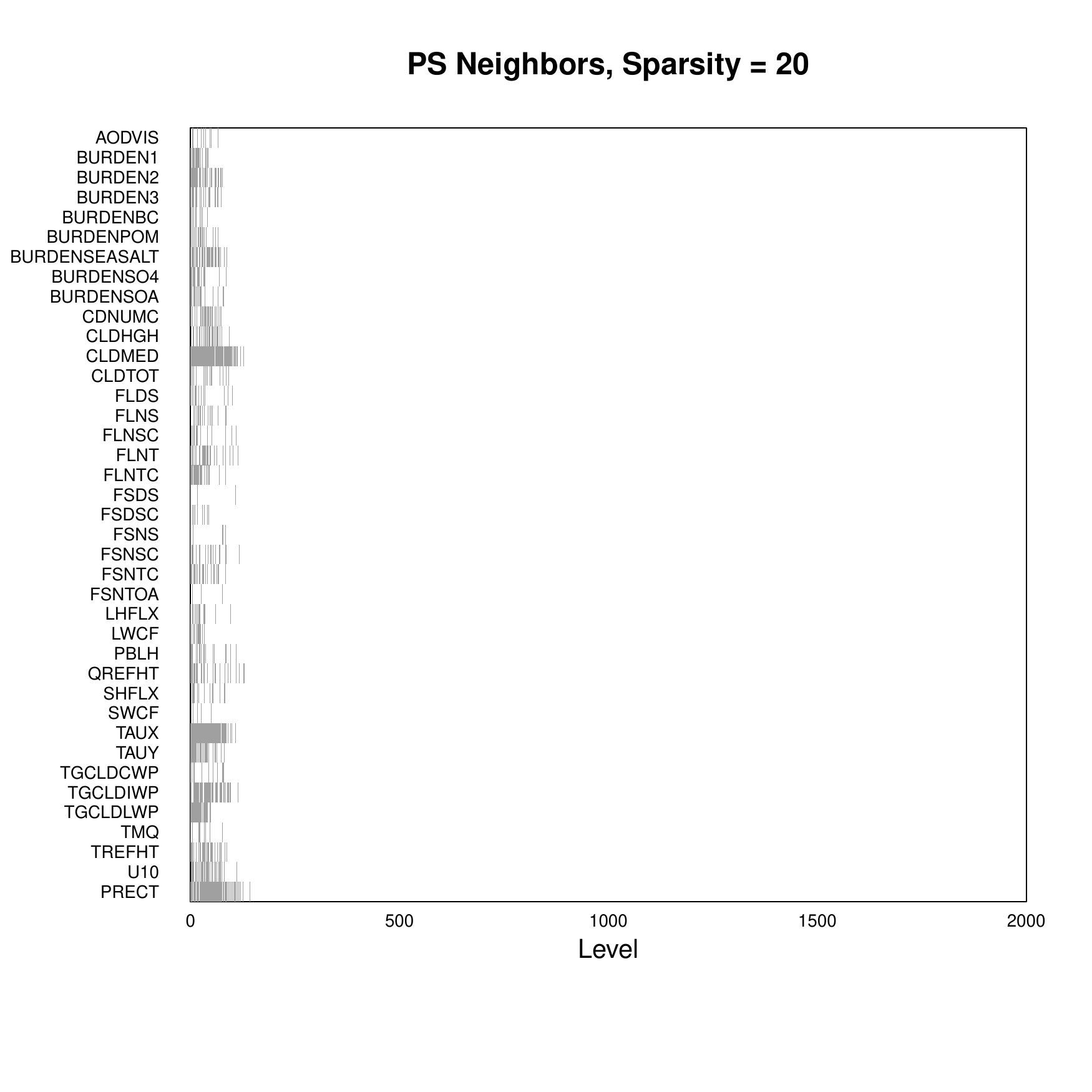} 
  \end{minipage} 
  
     \begin{minipage}[b]{0.32\linewidth}
    \centering
    \includegraphics[width=\linewidth]{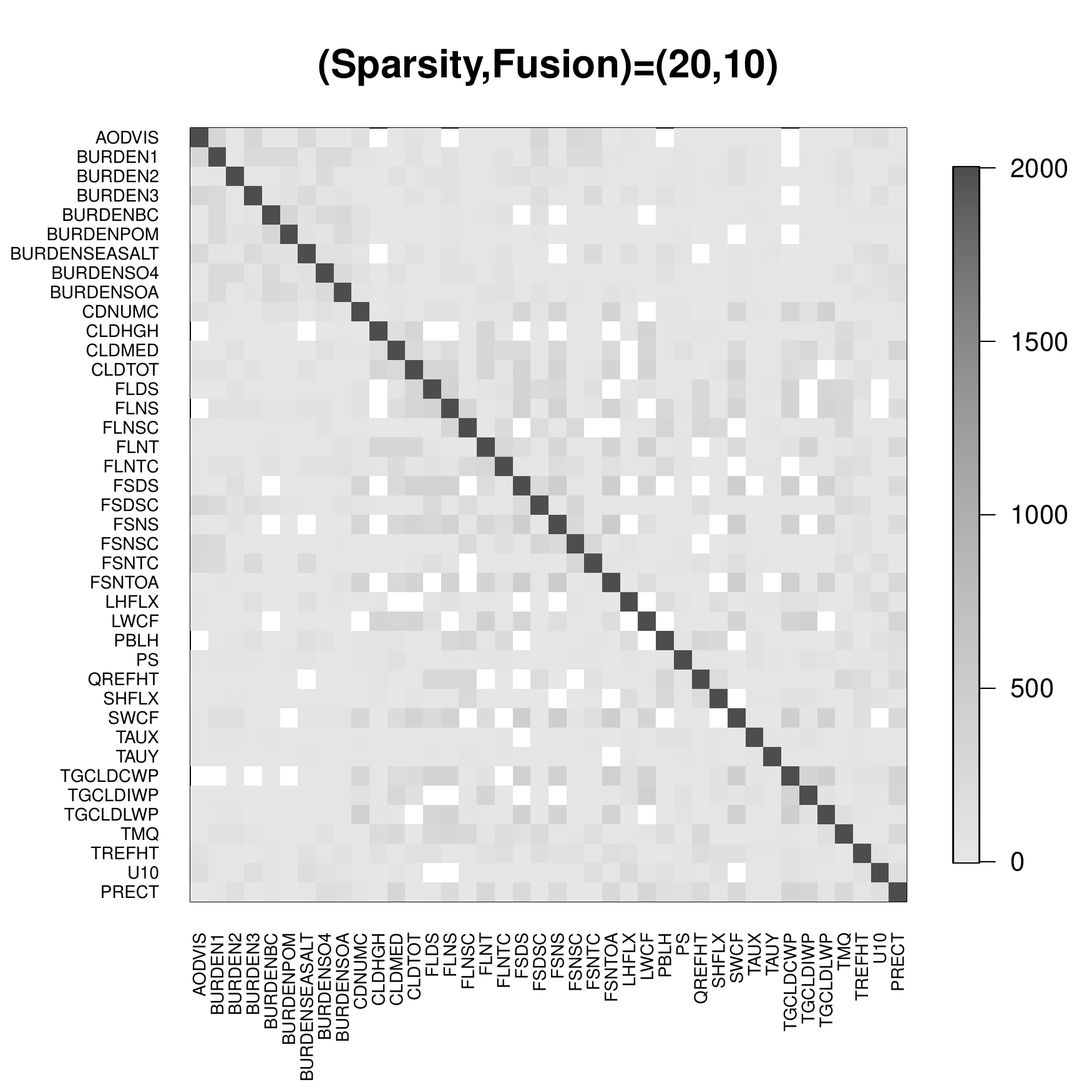} 
  \end{minipage} 
      \begin{minipage}[b]{0.32\linewidth}
    \centering
    \includegraphics[width=\linewidth]{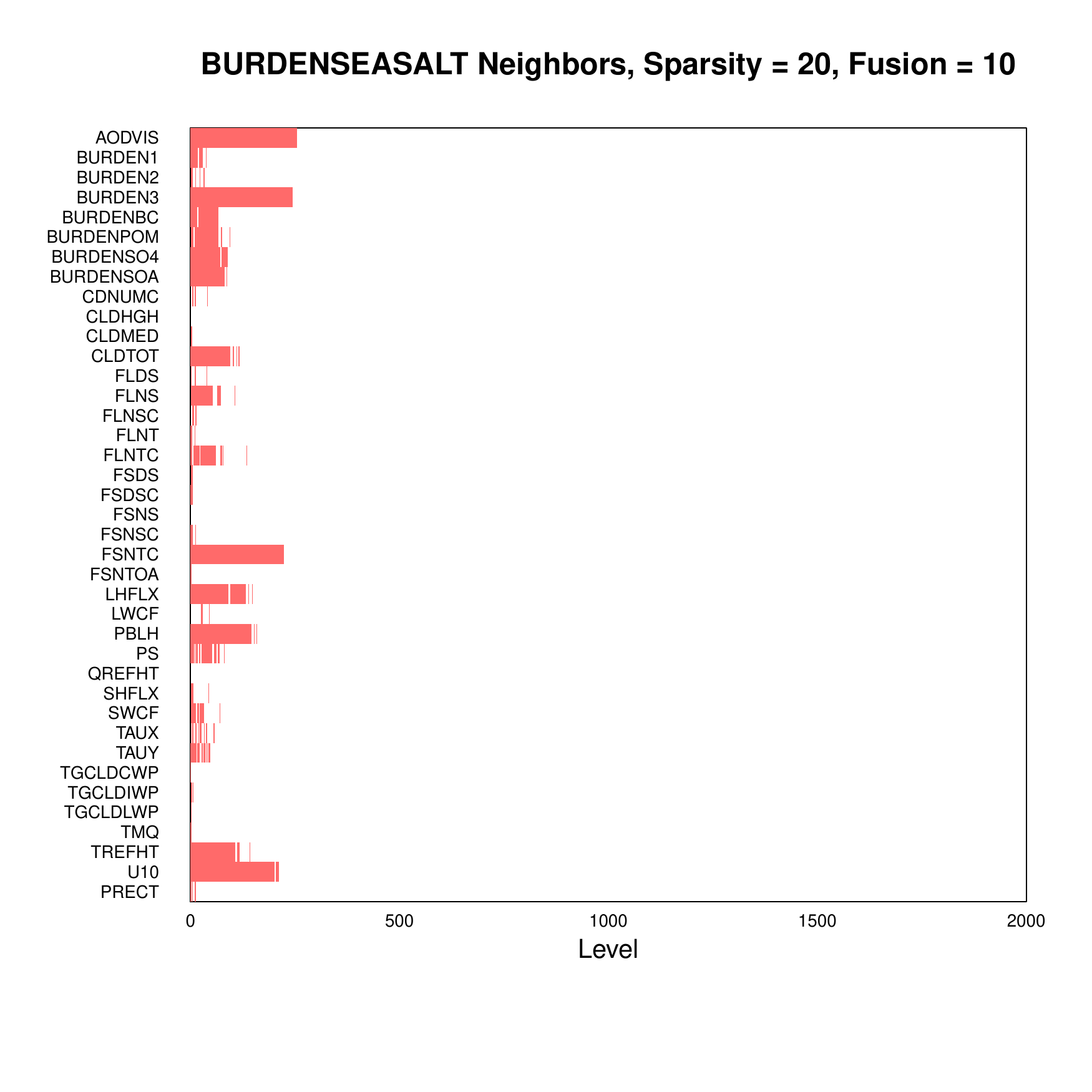} 
  \end{minipage} 
   \begin{minipage}[b]{0.32\linewidth}
    \centering
    \includegraphics[width=\linewidth]{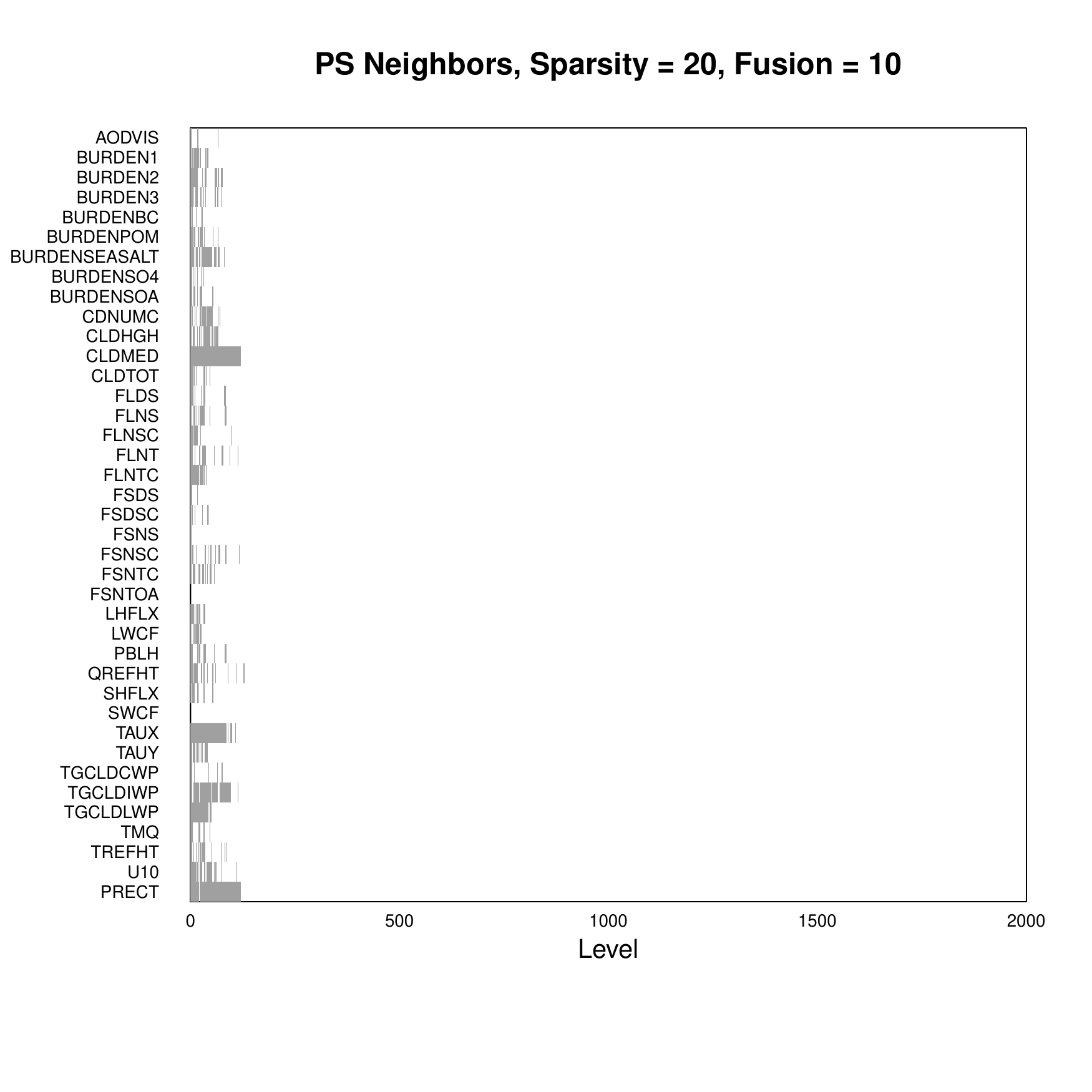} 
  \end{minipage} 
  \vspace{-.1in}
      \caption{Illustrating how graphical model neighborhoods behave for various penalty choices. Left column counts the nonzeros of $\hat{\mathbf{Q}}_1,\dots,\hat{\mathbf{Q}}_{2000}$ by level. Center and right columns show how the neighbors of BURDENSEASALT and PS change over level.}
  \label{Qneighborbylevel}

  \end{figure}

\begin{figure}[h!]
\centering
  \includegraphics[width=0.95\linewidth]{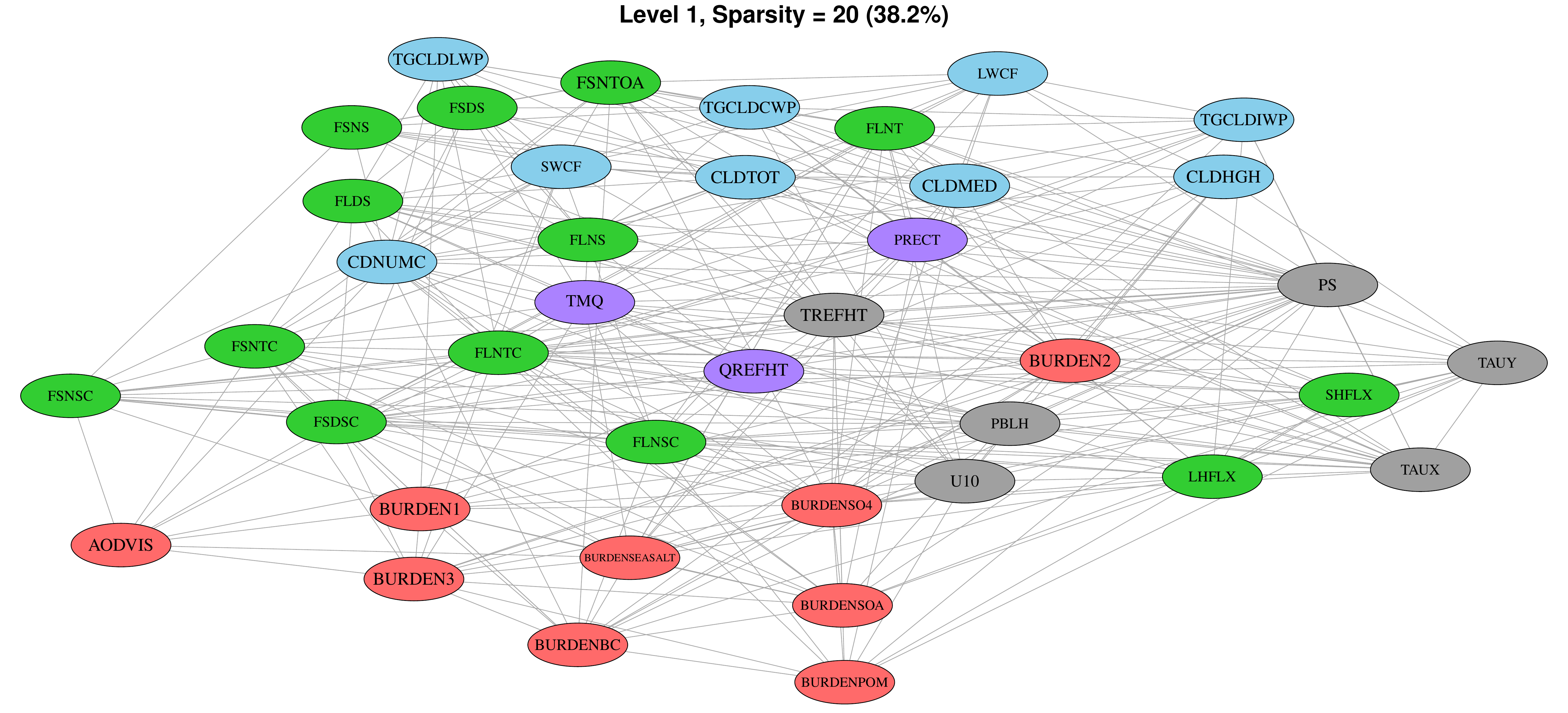}
  \includegraphics[width=0.95\linewidth]{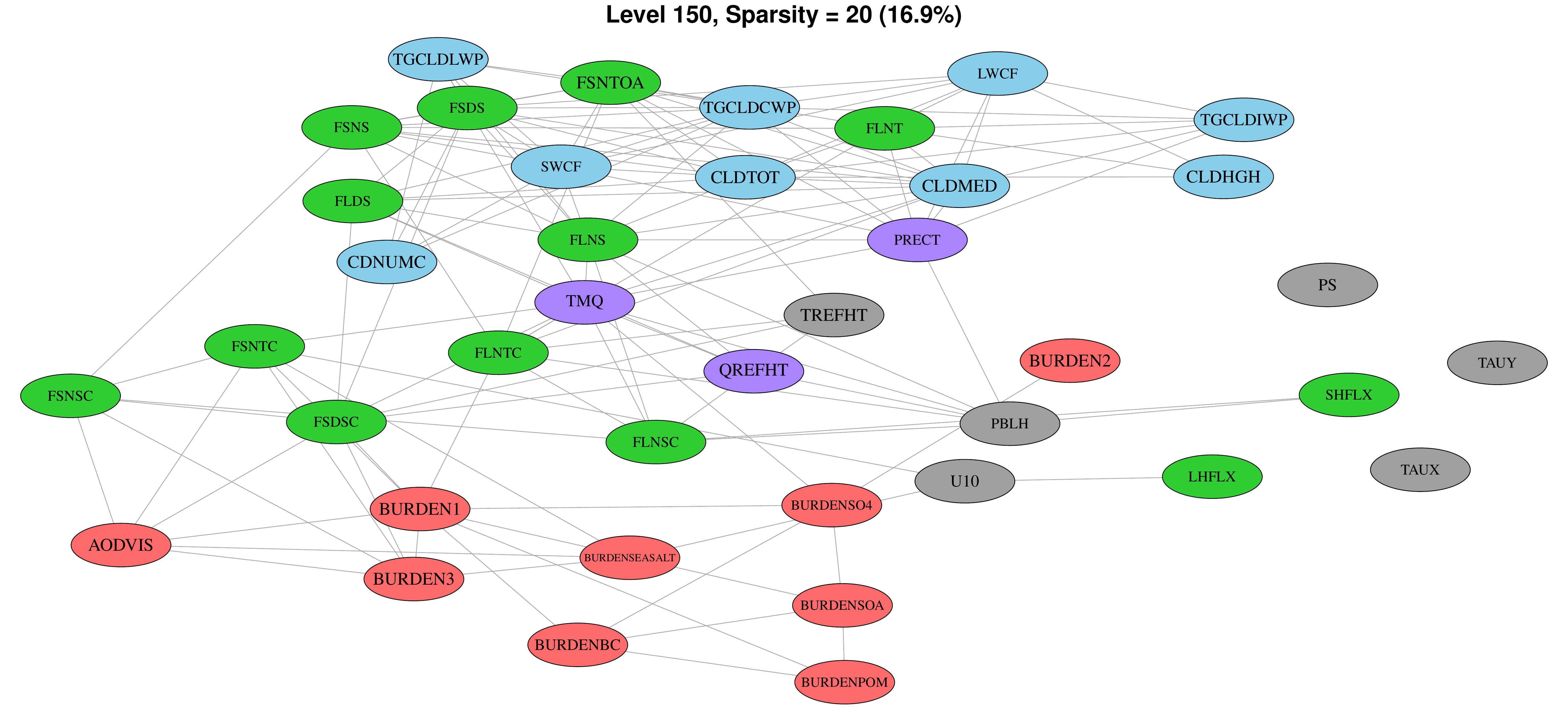}
  \includegraphics[width=0.95\linewidth]{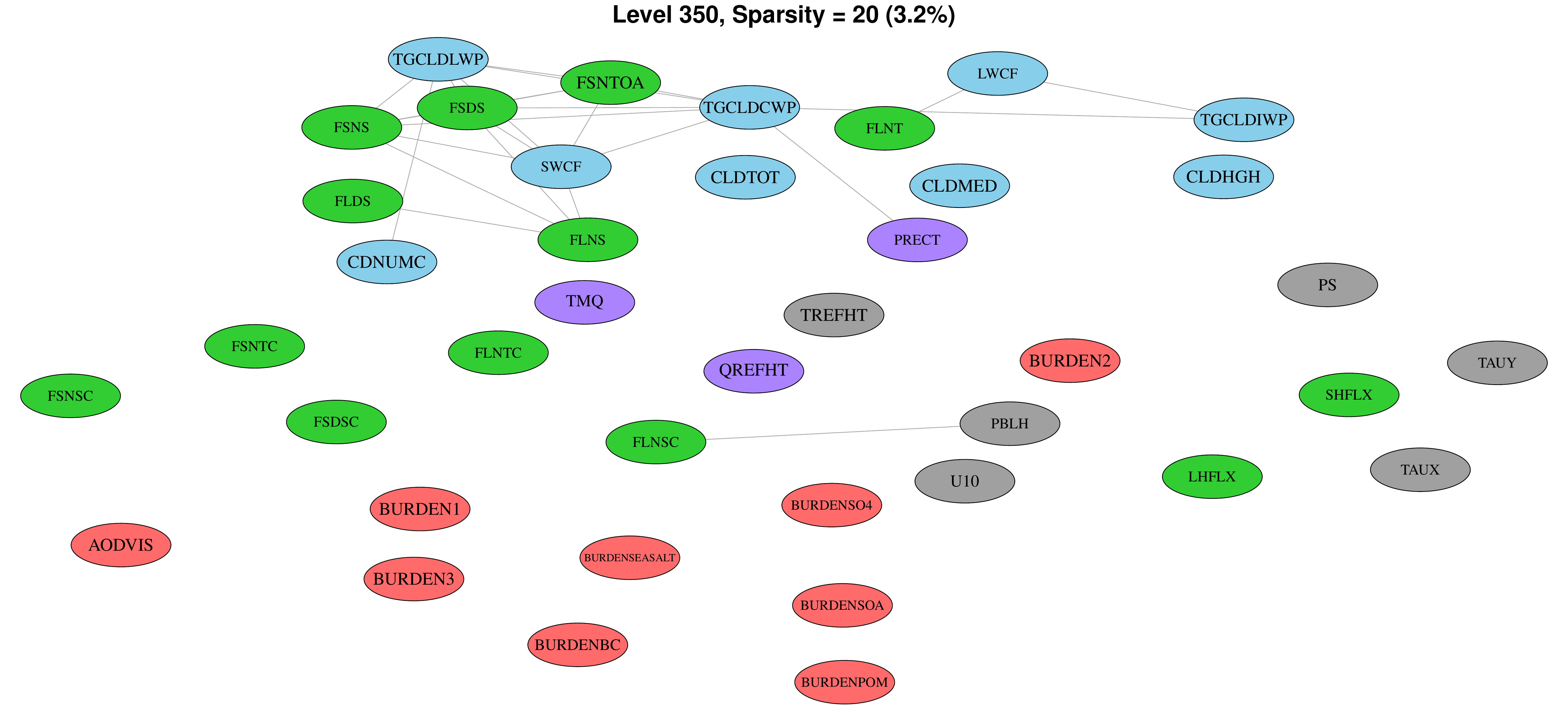}
        \caption{Example estimated graphical models for $\lambda=20$. Low level graphs are noisy, with the Level 1 graph containing 38.2\% of all possible connections. Higher level graphs show reasonable variable clusters until eventually no graph edges exist.}
\label{Qgraphs}

  \end{figure}

\begin{figure}[h!]
  \centering
      \includegraphics[width=.9\linewidth]{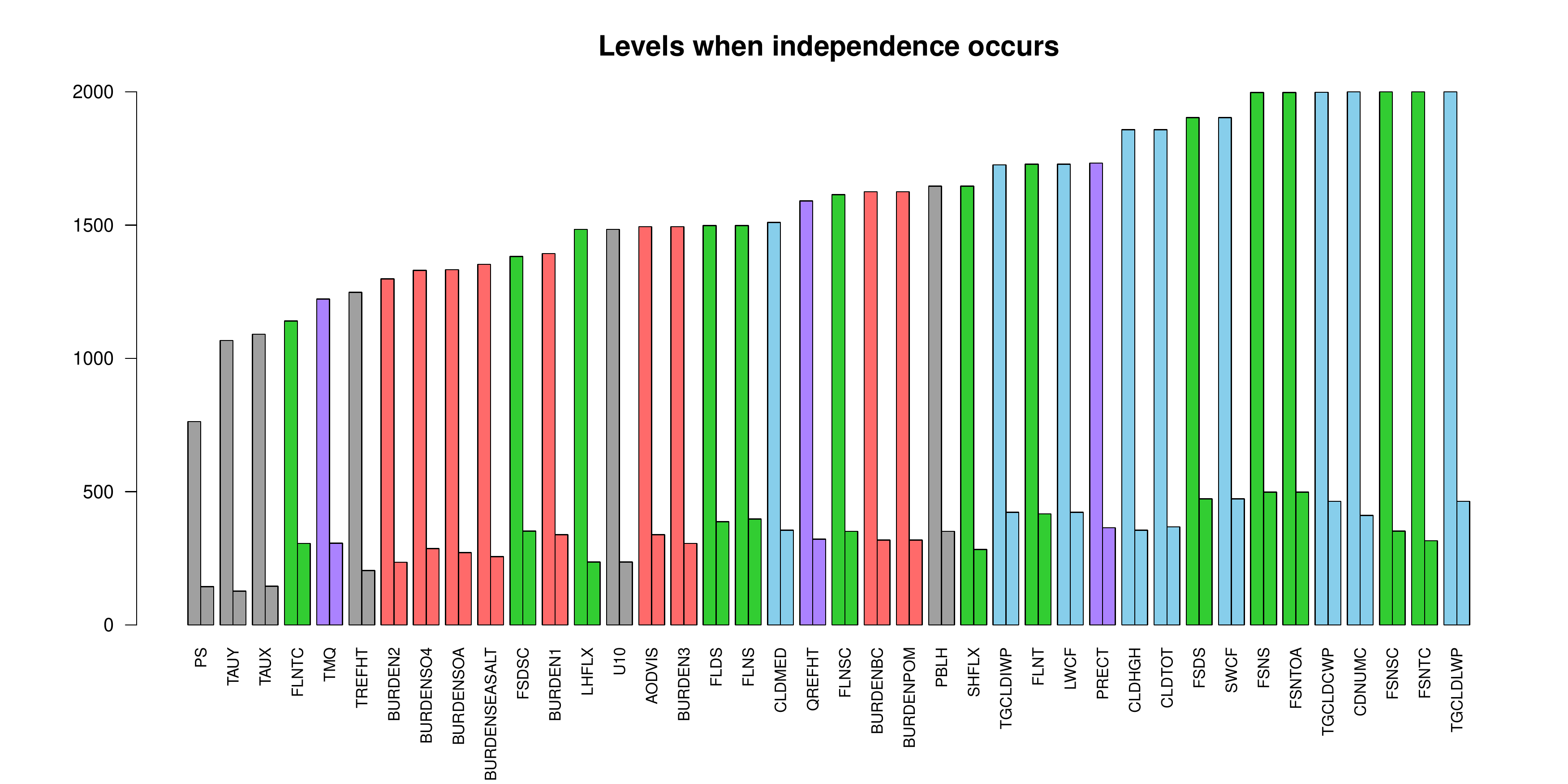}
    \caption{The first level at which a variable becomes (and remains) independent. Variables are ordered in increasing fashion according to the implied independence level for $\lambda=1$. Results for $\lambda=20$ accompany the taller bars and show a similar story for variable groups but with earlier levels of independence.}
\label{zeroout}
  \end{figure}

Our penalized maximum likelihood procedure produces interpretable graphical models which imply conditional independencies among the variables at varying levels of resolution. Here we study some properties of the estimated graphical models. Figure \ref{Qneighborbylevel} gives an idea of how the conditional independence structure changes with respect to level for different penalties. The left column counts the nonzeros of the estimated precision matrices $\hat{\mathbf{Q}}_1,\dots,\hat{\mathbf{Q}}_{2000}$ over level. With $\lambda=20$ we see several conditional independencies between variables that persist over all levels. The middle and right columns show how the conditional dependence neighborhood structure of two variables (BURDENSEASALT and PS) changes over level, and the impact of the fusion penalty is most apparent in these columns. Broadly speaking, the fusion penalty smooths out the neighbor pattern across levels. Note that the fusion penalty can fuse adjacent nonzeros rather than adjacent zeros and can cause a neighbor to fuse across all levels if $\rho$ is large enough. 
       Figure \ref{Qgraphs} displays estimated graphical models for different levels of resolution. At early levels corresponding to large scale spatial patterns, the graphs are quite dense. Higher frequency EOFs display interesting and reasonable patterns. For example, near the top of the $\ell=150$ graph we see groupings among many cloud and precipitation variables, and these connections are more evident at higher levels and even with different penalty parameters (not shown). For higher level EOFs, past around $\ell=500$ with $\lambda=20$, the graphs suggest variable independence. 
	This idea of complete independence is explored in Figure \ref{zeroout} where we display the first level at which each variable is independent and remains independent of all other variables. Note again a natural grouping of variable types, with transport and pressure variables achieving independence much earlier than finer-scale precipitation and cloud variables. Further, as could also be observed in the visualization of the conditional dependence structure for the two selected variables in Figure \ref{Qneighborbylevel}, the level at which independence occurs varies quite dramatically with the sparsity penalty, leading to roughly a four-fold increase in the number of connected levels going from $\lambda=20$ to $\lambda=1$.

\begin{figure}[h!]
\centering

  \begin{minipage}[b]{0.32\linewidth}
    \centering
    \includegraphics[width=\linewidth]{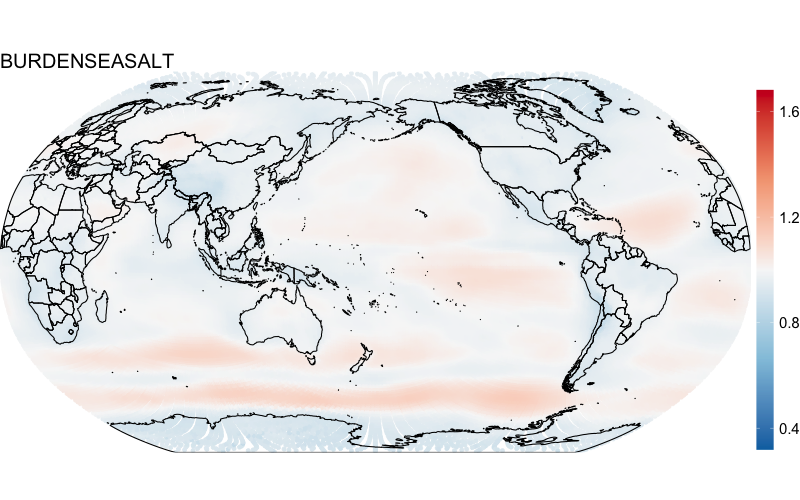} 
  \end{minipage} 
      \begin{minipage}[b]{0.32\linewidth}
    \centering
    \includegraphics[width=\linewidth]{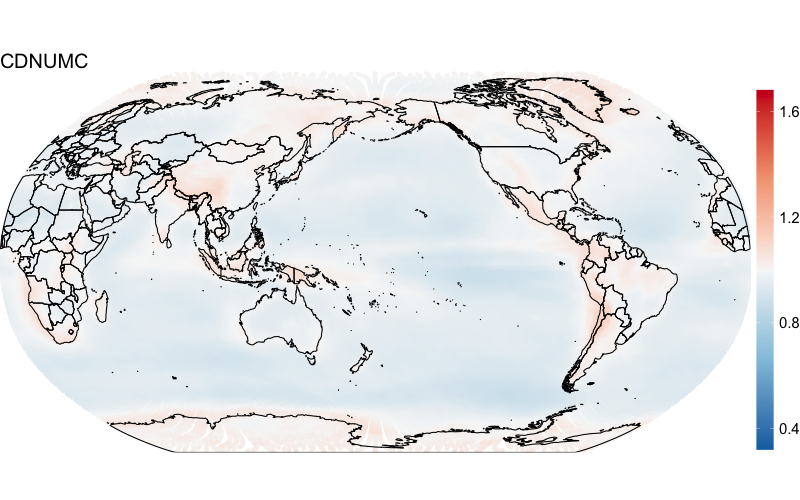} 
  \end{minipage} 
   \begin{minipage}[b]{0.32\linewidth}
    \centering
    \includegraphics[width=\linewidth]{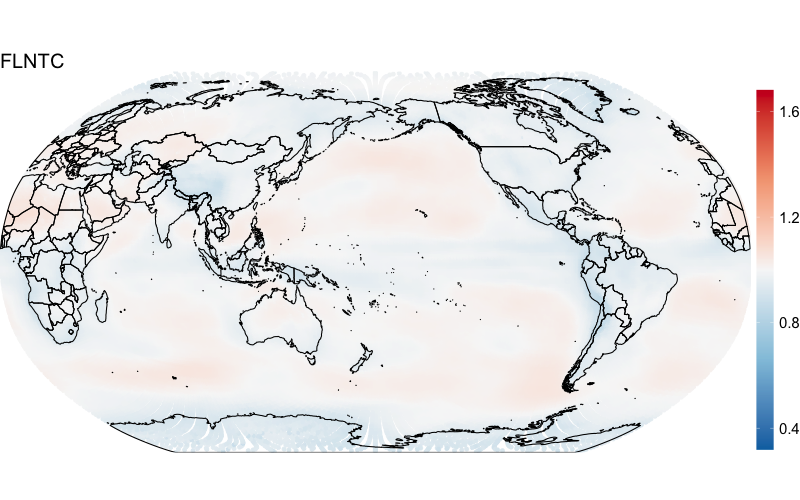} 
  \end{minipage} 
  
    \begin{minipage}[b]{0.32\linewidth}
    \centering
    \includegraphics[width=\linewidth]{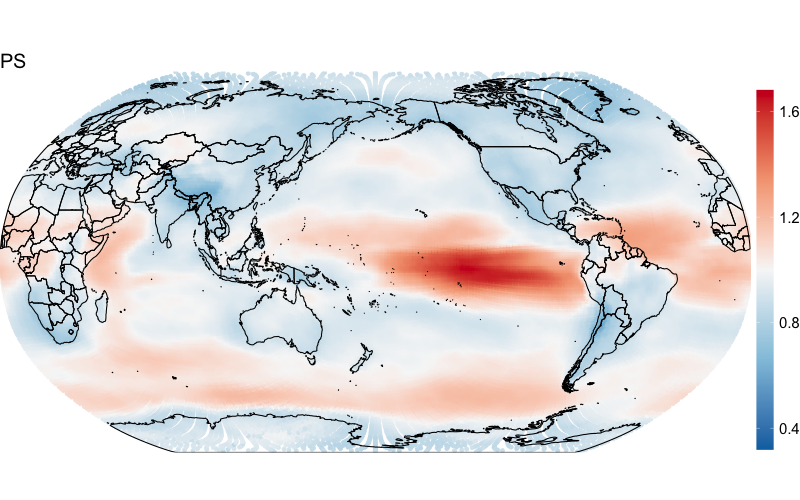} 
  \end{minipage} 
      \begin{minipage}[b]{0.32\linewidth}
    \centering
    \includegraphics[width=\linewidth]{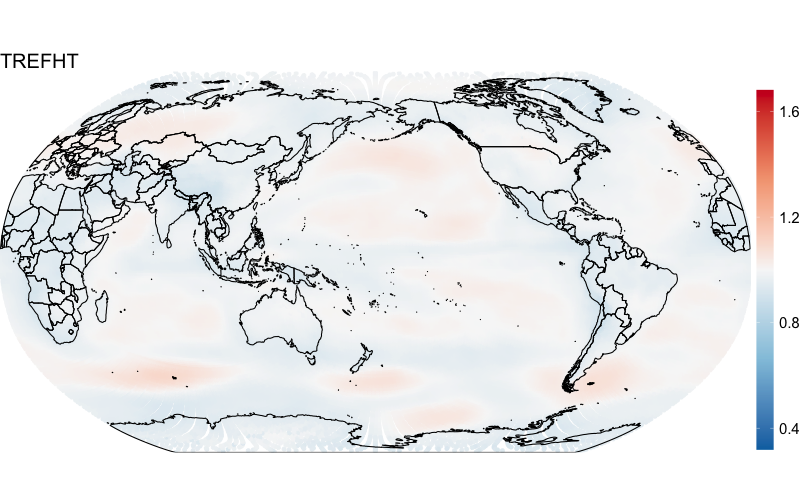} 
  \end{minipage} 
   \begin{minipage}[b]{0.32\linewidth}
    \centering
    \includegraphics[width=\linewidth]{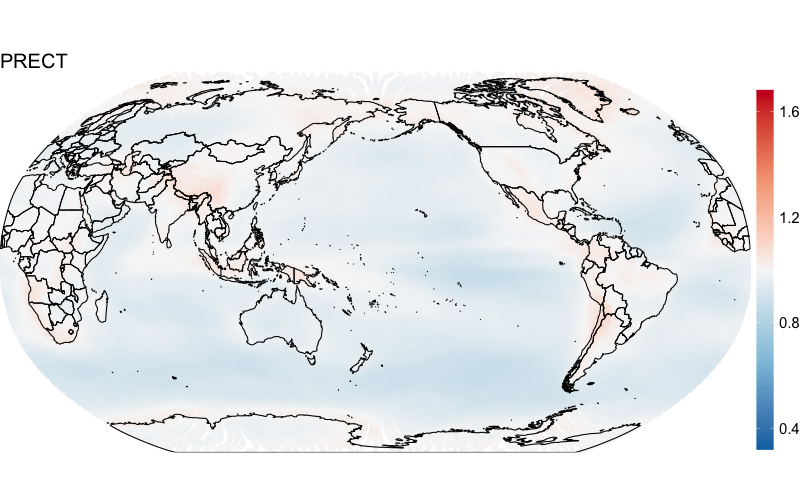} 
  \end{minipage}

    \caption{Estimated local standard deviations for six variables. Standardization of the variables is reflected in the color scale, where unit values are colored white. Values above one suggest the empirical standardization does not explain enough variability.}
        \label{Qsd}
\end{figure}

  Now, we examine some spatial properties of our estimates. Figure \ref{Qsd} shows the estimated local standard deviations for a subset of six variables. Recall that the variables were  standardized to have an empirical unit standard deviation, so these plots should be interpreted as potential bias corrections where the standardization fails to accurately describe the variability. Most striking is the El Ni\~no effect apparent in the plot for the pressure variable PS. The pattern's presence is unsurprising since El Ni\~no/La Ni\~na are strongly tied to changes in pressure over the Pacific Ocean, and their relative infrequency likely requires more modeling care than just an empirical standardization. Finer-scale variables CDNUMC and PRECT are able to capture distinct behavior in mountainous regions (e.g., Rocky Mountains and Himalayas).

\begin{figure}[h!] 

  \begin{minipage}[b]{0.32\linewidth}
    \centering
    \includegraphics[width=\linewidth]{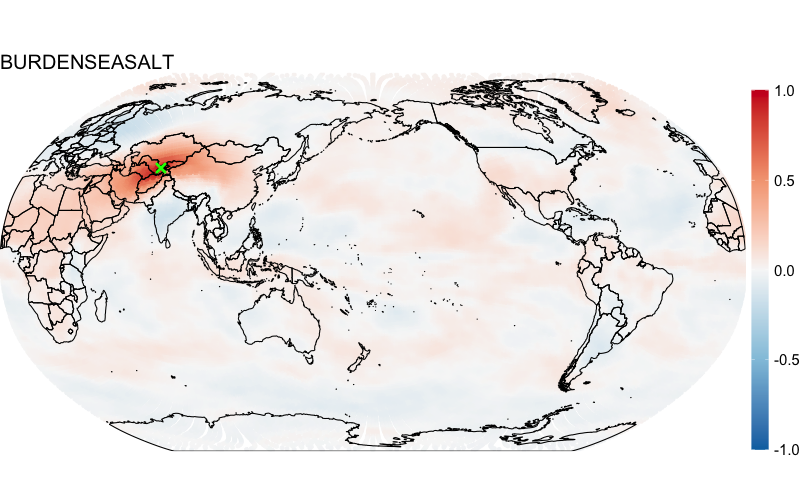} 
  \end{minipage} 
      \begin{minipage}[b]{0.32\linewidth}
    \centering
    \includegraphics[width=\linewidth]{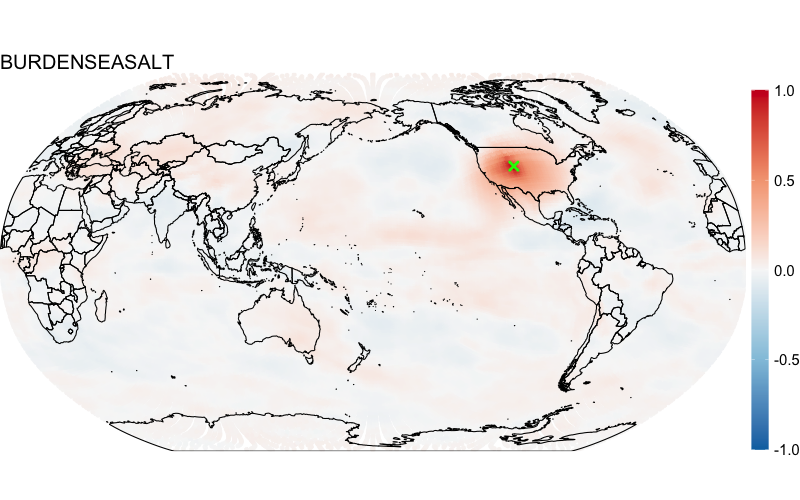} 
  \end{minipage} 
   \begin{minipage}[b]{0.32\linewidth}
    \centering
    \includegraphics[width=\linewidth]{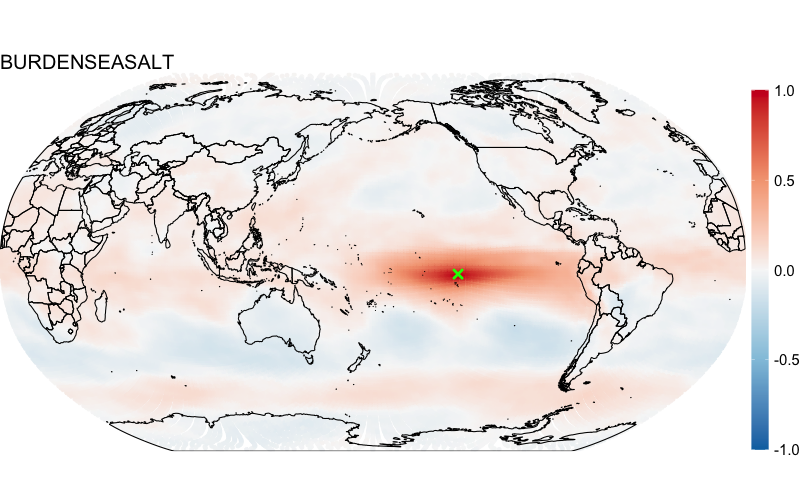} 
  \end{minipage}

    \begin{minipage}[b]{0.32\linewidth}
    \centering
    \includegraphics[width=\linewidth]{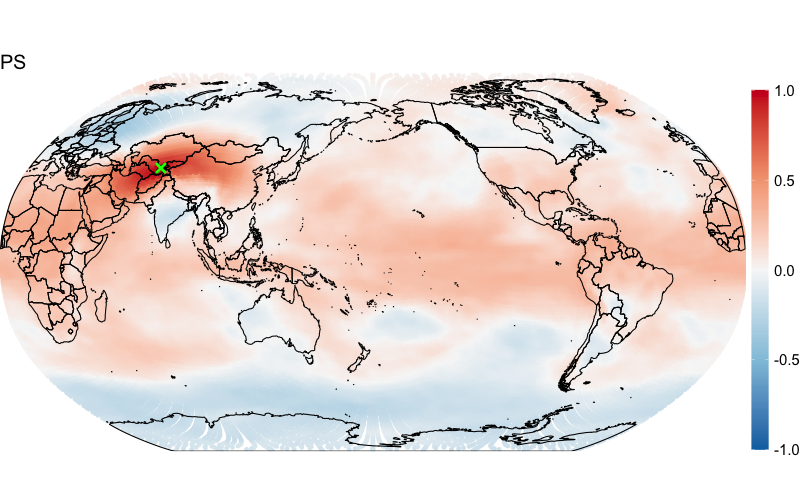} 
  \end{minipage} 
      \begin{minipage}[b]{0.32\linewidth}
    \centering
    \includegraphics[width=\linewidth]{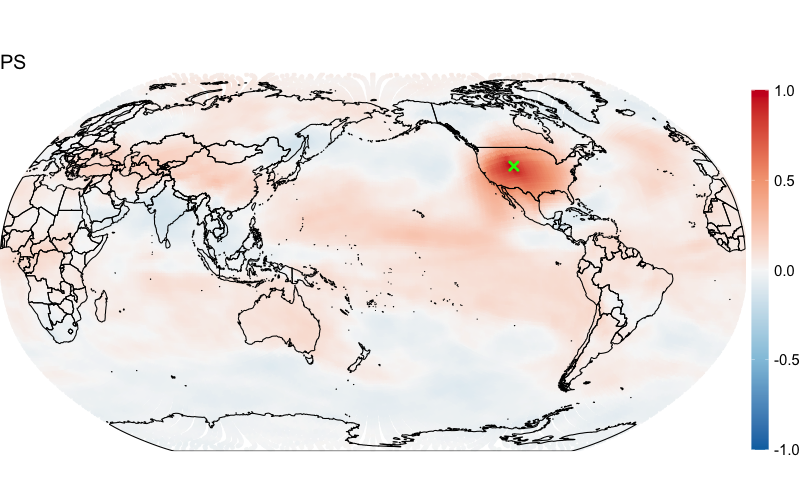} 
  \end{minipage} 
   \begin{minipage}[b]{0.32\linewidth}
    \centering
    \includegraphics[width=\linewidth]{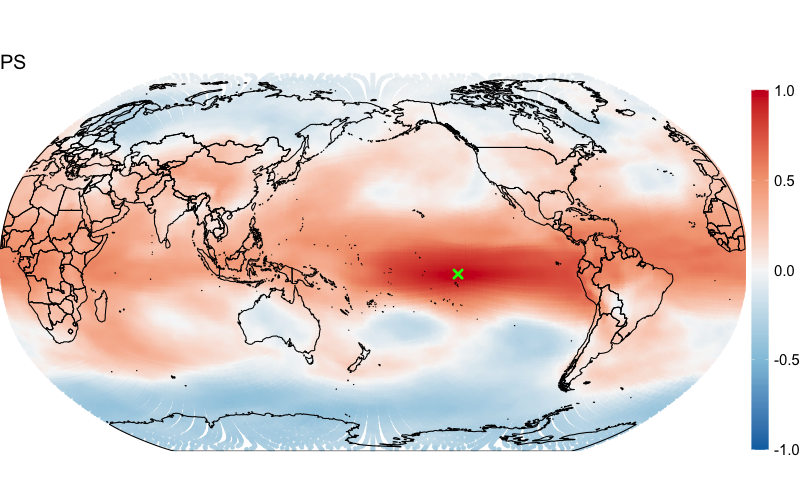} 
  \end{minipage}

      \begin{minipage}[b]{0.32\linewidth}
    \centering
    \includegraphics[width=\linewidth]{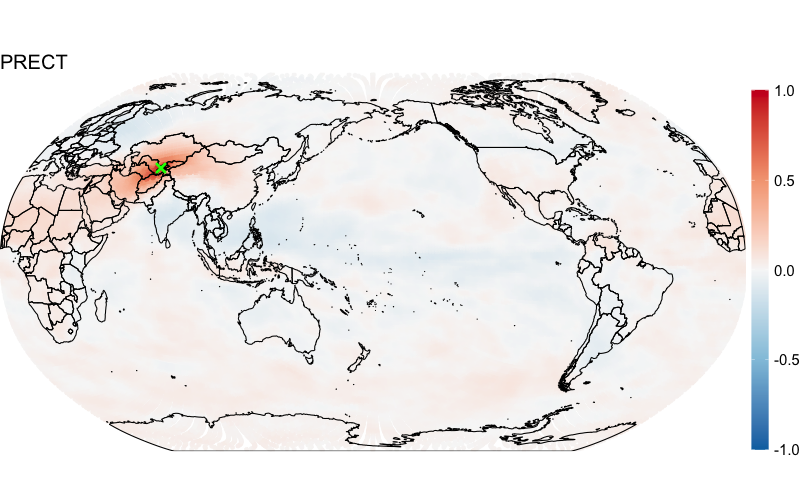} 
  \end{minipage} 
      \begin{minipage}[b]{0.32\linewidth}
    \centering
    \includegraphics[width=\linewidth]{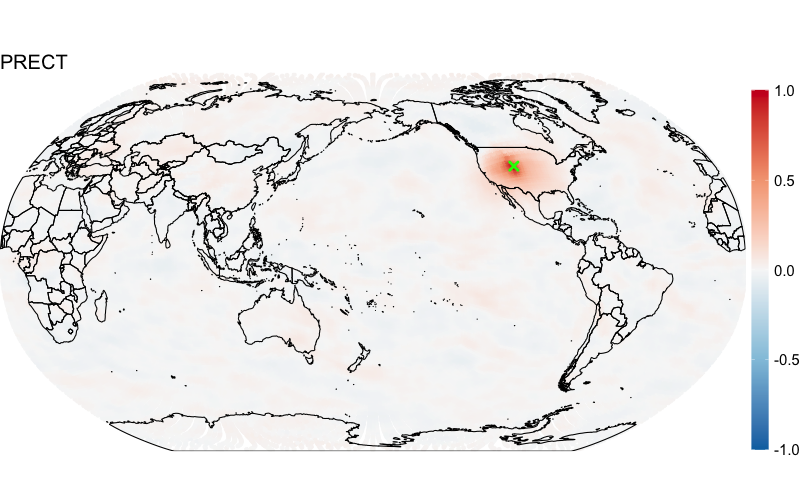} 
  \end{minipage} 
   \begin{minipage}[b]{0.32\linewidth}
    \centering
    \includegraphics[width=\linewidth]{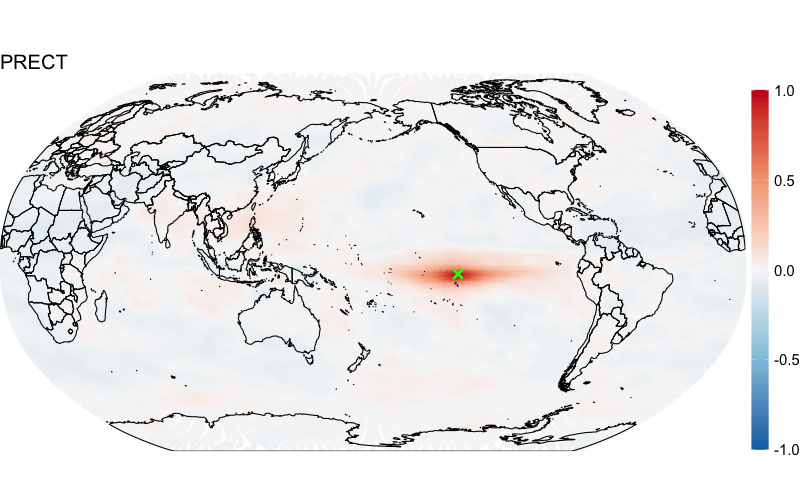} 
  \end{minipage}

    \caption{Estimated spatial correlation functions for BURDENSEASALT (top row), PS (middle row), and PRECT (bottom row). Correlation function is centered over Tajikstan in the left column, U.S.\ in the middle column, and the Pacific Ocean in the right column.}
  \label{Qspatialcor}

  \end{figure}

\begin{figure}[h!]

  \begin{minipage}[b]{0.32\linewidth}
    \centering
    \includegraphics[width=\linewidth]{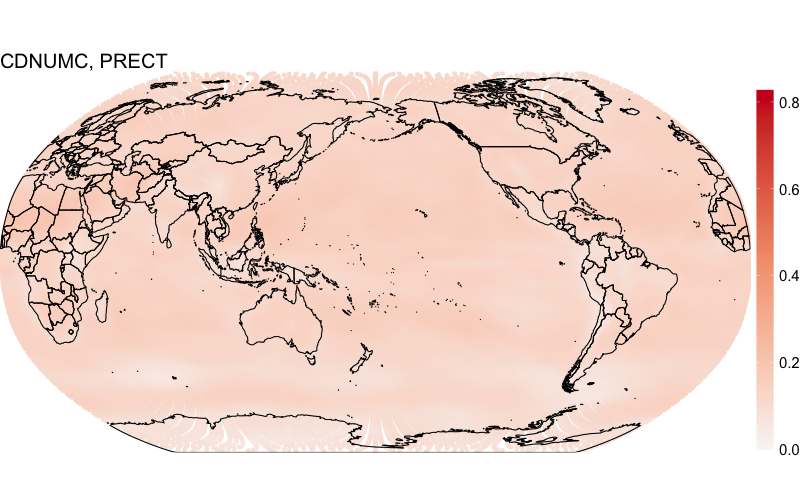} 
  \end{minipage} 
      \begin{minipage}[b]{0.32\linewidth}
    \centering
    \includegraphics[width=\linewidth]{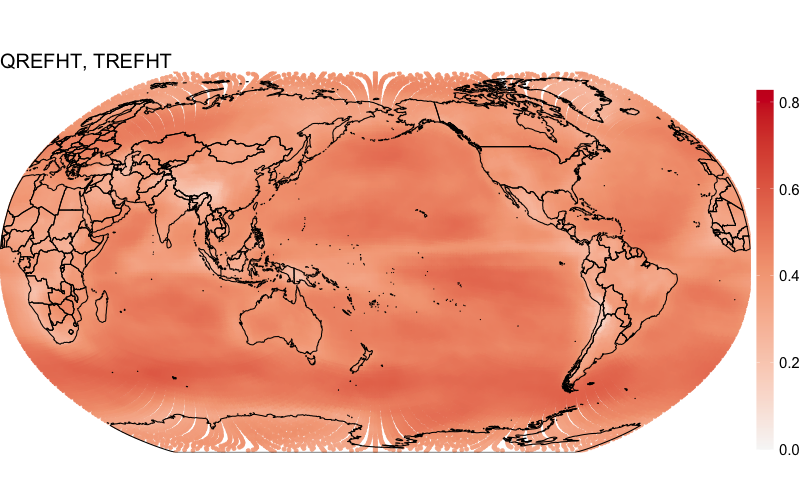} 
  \end{minipage} 
   \begin{minipage}[b]{0.32\linewidth}
    \centering
    \includegraphics[width=\linewidth]{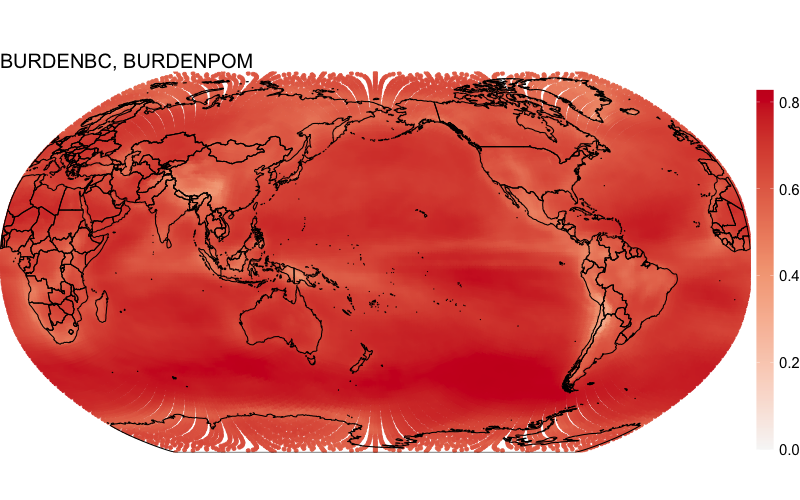} 
  \end{minipage} 
  
    \begin{minipage}[b]{0.32\linewidth}
    \centering
    \includegraphics[width=\linewidth]{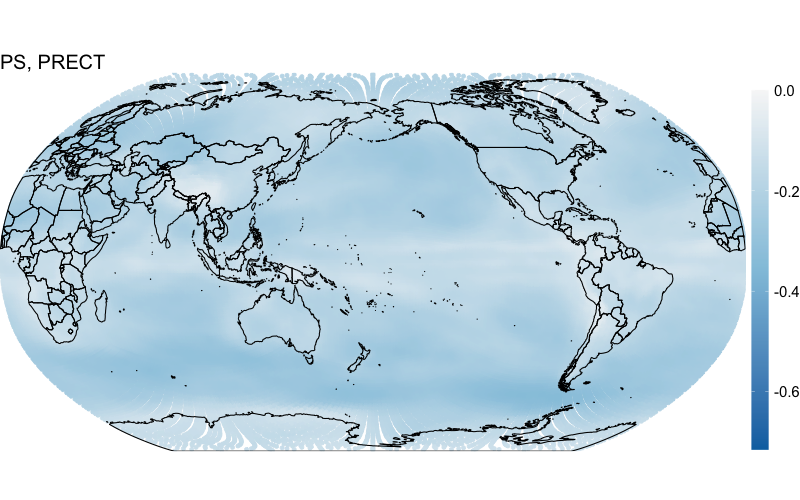} 
  \end{minipage} 
      \begin{minipage}[b]{0.32\linewidth}
    \centering
    \includegraphics[width=\linewidth]{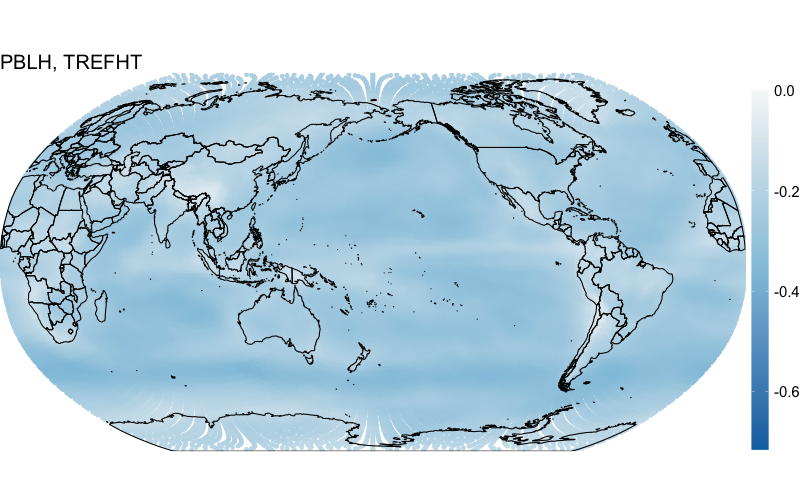} 
  \end{minipage} 
   \begin{minipage}[b]{0.32\linewidth}
    \centering
    \includegraphics[width=\linewidth]{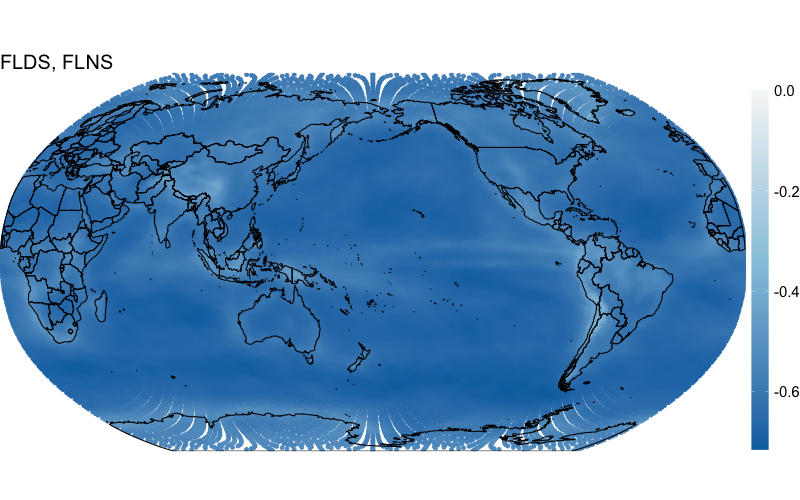} 
  \end{minipage} 
  
    \caption{Estimated local cross-correlations for several pairs of variables. Color scale changes between rows to permit easier comparisons between positively correlated variables (top row) and between negatively correlated variables (bottom row).}
  \label{Qpixelwisecrosscor}
\end{figure}

\begin{figure}[h!] 

  \begin{minipage}[b]{0.32\linewidth}
    \centering
    \includegraphics[width=\linewidth]{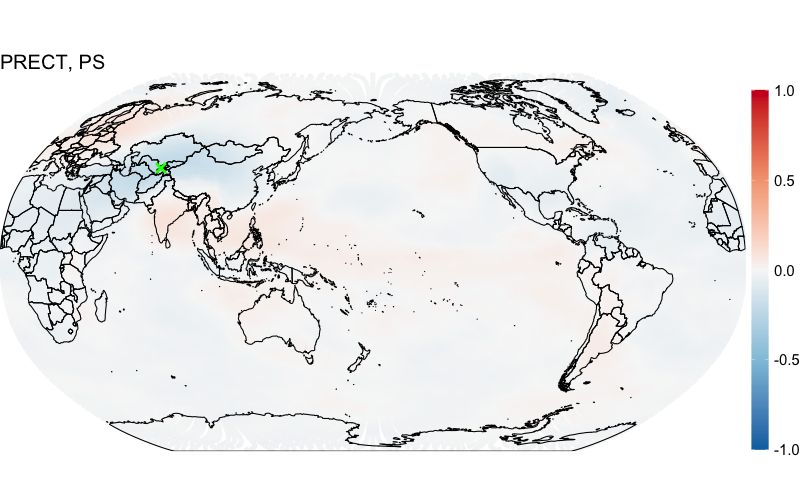} 
  \end{minipage} 
      \begin{minipage}[b]{0.32\linewidth}
    \centering
    \includegraphics[width=\linewidth]{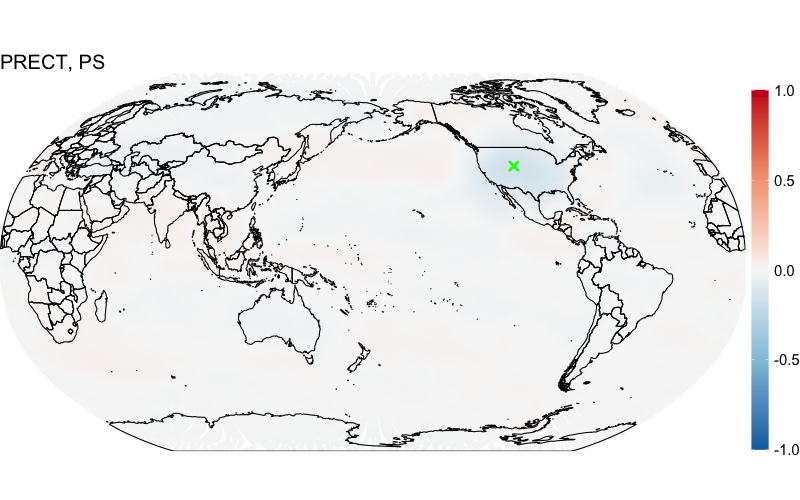} 
  \end{minipage} 
   \begin{minipage}[b]{0.32\linewidth}
    \centering
    \includegraphics[width=\linewidth]{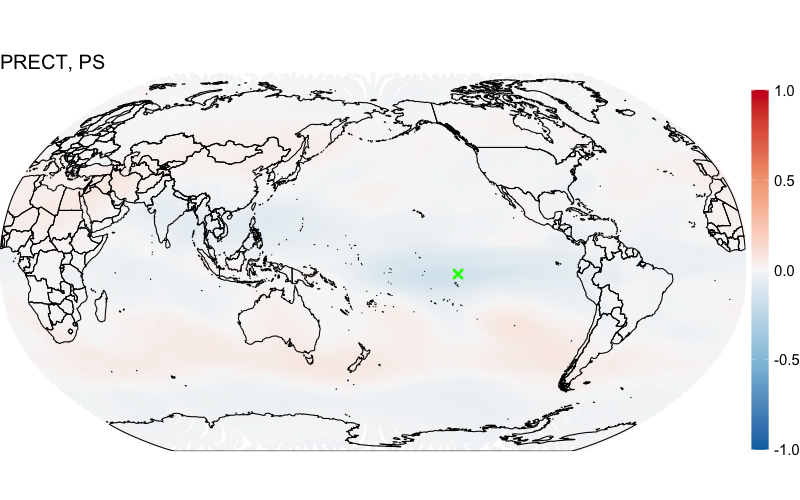} 
  \end{minipage}   
  
      \begin{minipage}[b]{0.32\linewidth}
    \centering
    \includegraphics[width=\linewidth]{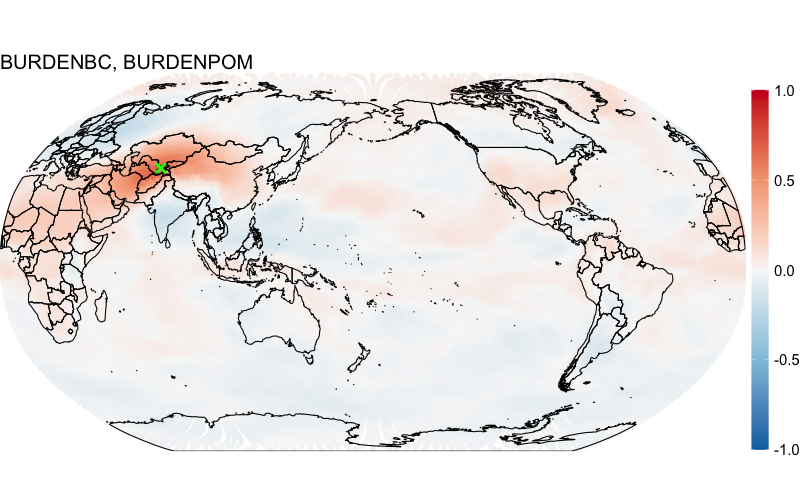} 
  \end{minipage} 
      \begin{minipage}[b]{0.32\linewidth}
    \centering
    \includegraphics[width=\linewidth]{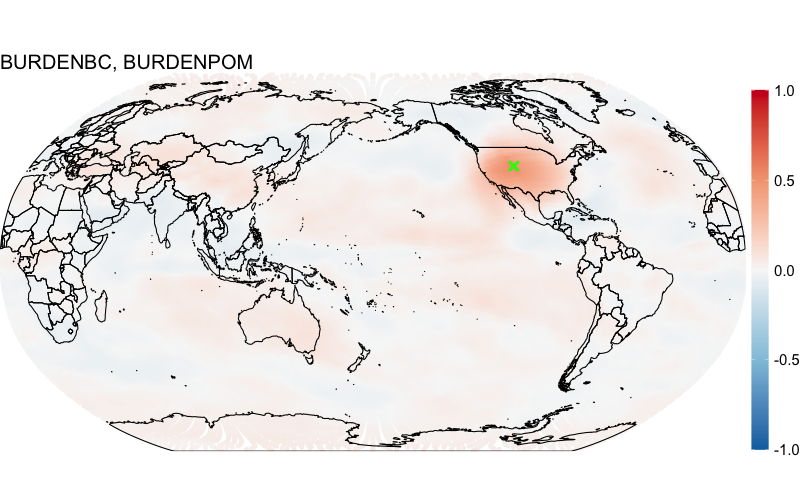} 
  \end{minipage} 
   \begin{minipage}[b]{0.32\linewidth}
    \centering
    \includegraphics[width=\linewidth]{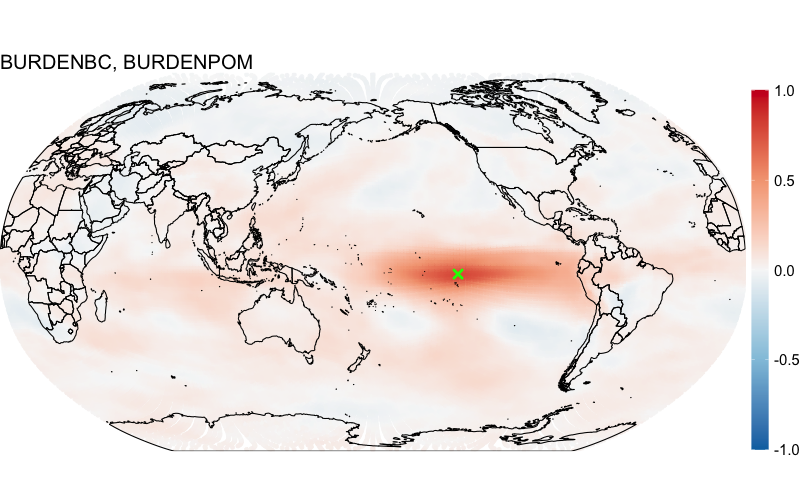} 
  \end{minipage}

    \caption{Estimated spatial cross-correlation functions, again centered over Tajikstan in the left column, U.S.\ in the middle column, and the Pacific Ocean in the right column. Top row shows PRECT and PS which exhibit negative local cross-correlations, while bottom row shows BURDENBC and BURDENPOM which exhibit positive local cross-correlations.}
  \label{Qspatialcrosscor}

  \end{figure}

In our last collection of figures, we examine correlation over space as well as across variables. Figure \ref{Qspatialcor} shows variable correlations as a function of space. To be precise, each image shows the correlation between the location marked in green and all other locations. Nonstationarity is evident from the difference in behaviors between land and ocean, and long range negative correlation is seen to be a possible byproduct of this modeling scheme.
In Figure \ref{Qpixelwisecrosscor}, we display estimated local cross-correlations between a few of our selected variables. Expected negative and positive correlations between pairs of variables are correctly captured (e.g., between pressure and precipitation and between cloud droplet concentration and precipitation).  
Finally, in Figure \ref{Qspatialcrosscor}, we show variable cross-correlations as a function of space. To be precise, each image shows the cross-correlation between the first variable at the location marked in green and the second variable at all other locations. The flexibility of the model is again apparent in its nonstationary behavior and various positive and negative cross-correlations. Such behavior is difficult to accommodate using extant models but readily available in our approach without any additional effort \citep{kleiber2013biom}. 
Both the correlation and cross-correlation functions centered where El Ni\~no/La Ni\~na occur (see right column in Figures \ref{Qspatialcor} and \ref{Qspatialcrosscor}) exhibit long-range dependence through the equator across the Pacific Ocean.

We conclude the data analysis with a brief commentary about the timing results, assuming the orthogonal basis has been constructed and $\tau_1^2,\dots,\tau_p^2$ are estimated. Note that the latter step is fairly quick using the technique in Section \ref{errorvariance_estimate} with an orthogonal basis. The choice of penalty dictates the runtime of the DC algorithm. For the maximum likelihood estimate with no penalty, the tolerance $\epsilon = 0.05$ is reached in fifteen DC iterations in five seconds. With sparsity penalty $\lambda=20$, the algorithm converges in two DC iterations in forty-five seconds. With sparsity penalty  $\lambda=1$, the algorithm again requires two DC iterations but takes thirteen minutes. With fusion penalty $\rho=10$, both estimates converged in one DC iteration using the $\rho=0$ solution as the initial guess, but $(\lambda,\rho)=(20,10)$ took seven minutes while $(\lambda,\rho) =  (1,10)$ took one minute.  All experiments were performed in Matlab on a MacBook Pro with a 6 core 2.6 GHz Intel Core i7 processor and 32 GB of RAM.

\clearpage

\section{Conclusion}
\label{conclusions}

Models for high-dimensional multivariate spatial processes struggle to accommodate nonstationary data with a large number of variables and observation locations. 
We have presented a multivariate Gaussian process model which will be applicable in a variety of future endeavors. There are many benefits under this framework, including nonstationary covariance and cross-covariance functions, exact likelihood calculations, and cheap computations and storage when $\mathbf{W}_1,\dots,\mathbf{W}_L$ are independent and basis functions are orthogonal.

Future experiments could relax the assumption of independent weights and estimate the entire sparse $pL \times pL$ precision matrix $\mathbf{Q}$ as in Section \ref{BGLmethodology} with the DC algorithm (\ref{QUICproblem}). Even if the basis functions are not orthogonal, $ \boldsymbol \Phi^{\mathrm{T}} \mathbf D^{-1} \boldsymbol \Phi  = \Phi^{\mathrm{T}} \Phi \otimes \text{diag}(\tau_1^{-2},\dots,\tau_p^{-2})$ is sparse due to the Kronecker product with a diagonal matrix. In this formulation, it is crucial to
rewrite (\ref{linearization}) and calculate this linearization term by solving linear systems with the sparse matrix $\mathbf{Q} + \boldsymbol \Phi^{\mathrm{T}} \mathbf D^{-1} \boldsymbol \Phi $ or its sparse Cholesky decomposition. Note that the dense $pL \times pL$ linearization matrix must then be stored in memory, and advanced graphical lasso algorithms \citep{lineartime2019} must be explored for the subsequent graph estimation.
Modeling dependence across basis functions would allow for more flexibility in the cross-covariances, since, with the independence assumption, $\mathrm{Cov}(Z_i(\mathbf{s}),Z_j(\mathbf{s}')) = \sum_{\ell=1}^L \phi_\ell(\mathbf{s}) ((\mathbf{Q}_\ell)^{-1})_{ij} \phi_\ell(\mathbf{s}') = \mathrm{Cov}(Z_i(\mathbf{s}'),Z_j(\mathbf{s}))$ is symmetric.

We demonstrated that our model can easily fit a large climate ensemble and produce reasonable and interpretable results. Our method also easily scales to computer experiments with high-dimensional inputs (e.g.,\ $\mathbf{s} \in \mathbb{R}^d$ with $d \gg 0$). Extending these multivariate basis graphical lasso ideas to a space-time setting could further prove to be fruitful.

\section*{Acknowledgements}

This research was funded by grants NSF DMS-1821074 and NSF DMS-1923062.

  \clearpage

\bibliographystyle{rss} 
\bibliography{multivarbib}

\end{document}